\documentclass[12pt]{article}
\usepackage{amsbsy}

\let\vp=\varphi

\def \HH {{\rm H}}

\def \bn {{\bf n}}

\def \del {\partial}
\def \tg {\tilde \gamma}

\def \nfty {\infty}
\def \s {\sigma}
\def \ov {\over}
\def \td {\tilde}

\def \M {{\cal M}}
\def \MM {{\cal M}}

\def\K{{(2\pi\kappa_d L)}}
\def\K{{K}}

\newcommand{\be}{\begin{equation}}
\newcommand{\ee}{\end{equation}}
\newcommand{\bea}{\begin{eqnarray}}
\newcommand{\eea}{\end{eqnarray}}

\def \la{\label}
\newcommand{\rf}[1]{(\ref{#1})}

\font\mybb=msbm10 at 12pt

\def\bb#1{\hbox{\mybb#1}}

\def\Re {\bb{R}}

\def\ov{\over}

\def\id{\protect{{1 \kern-.28em {\rm l}}}}

\def\bfsigma{{\boldsymbol{\sigma}}}

\def \ha {{1\ov 2}}

\def\pint{{-\!\!\!\!\!\!\int}}

\def \adss{$AdS_5 \times S^5$\ }
\def\LM{Lunin-Maldacena }
\def\J{{\cal J}}
\def\N{{\cal N}}
\def\a{\alpha}
\def\b{\beta}
\def\tb{\tilde{\beta}}
\def\l{\lambda}
\def\tl{\tilde{\l}}
\def\z{\zeta}
\def\s{\sigma}
\def\ts{\tilde{\sigma}}

\def\g{\gamma}
\def\tg{\tilde{\gamma}}
\def\e{\epsilon}
\def\la{\label}
\def\={\, =\, }
\def\+{\, +\, }
\def\-{\, -\, }

\def\r{\rho}
\def\k{\kappa}

\def\={\, =\, }
\def\+{\, +\, }
\def\-{\, -\, }
\def\LL{{\mathcal L }}
\def\RR{{\mathcal R }}
\def \A {{\mathcal A}}
\def\p{\phi}
\def\ttp{\tilde{\tilde{\phi}}}

\def \pa{\partial}
\def \Tr {{\rm Tr}}
\def \T {{\rm T}}
\def\hg{\tilde{\gamma}}

\def \H{{\cal H}}

\def \bs {{\bf \sigma}}
\def \sql {\sqrt{\lambda}}\def \E {{\cal E}}

\def \YM {{\rm YM}}

\def \mm {{\rm m}}

\def \N {{\cal N}}
\def \ci{\cite}
\def \sym {$\N=4$ SYM \ }

\def \adss {${AdS}_5 \times {S}^5$\ }
\def \ci{\cite}
\def \foot {\footnote}
\def \N {{\cal N}}
\def \l {\lambda}

\def \S {section\ }

\def \bq {{\bf q}}

\def \bi {\bibitem}

\def \w {{\rm w}}

\def \th {\theta}\def \r {\rho}

\def \four {{\textstyle{1 \ov 4}}}
\def \ha {{\textstyle{1 \ov 2}}}

\def \bs  {\bfsigma}

\makeatletter
\renewcommand\section{\@startsection {section}{1}{\z@}%
                                   {-3.5ex \@plus -1ex \@minus -.2ex}%
                                   {2.3ex \@plus.2ex}%
                                   {\normalfont\large\bfseries}}

\renewcommand\subsection{\@startsection{subsection}{2}{\z@}%
                                   {-3.25ex\@plus -1ex \@minus -.2ex}%
                                   {1.5ex \@plus .2ex}%
       {\normalfont\normalsize\bfseries}}
\makeatother

\begin{document}


\begin{titlepage}

\begin{flushleft}
\hfill PUPT-2155
\end{flushleft}
\vspace{-1cm}

\begin{flushleft}
\end{flushleft}


\bigskip

\vspace{1cm}
\begin{center}
{\Large\bf
Gauge-string duality for superconformal deformations
\vspace{0.2cm}
of N=4 Super Yang-Mills theory
}

\vspace{0.3cm}

\vspace{.5cm} {S.A.~Frolov$^{a,}$\footnote{Also at Steklov
Mathematical Institute, Moscow. frolovs@sunyit.edu},
R.~Roiban$^{b,}$\footnote{rroiban@princeton.edu} and
A.A.~Tseytlin$^{c,}\footnote{Also at Imperial College London
and  Lebedev  Institute, Moscow. tseytlin@mps.ohio-state.edu}$}\\
\vskip 0.3cm

{\em $^{a}$Department of Mathematics/Sciences,
SUNY IT,\\
P.O. Box 3050, Utica, NY 13504-3050, USA\\
\vskip 0.1cm
$^{b}$Department of Physics, Princeton University,\\
 Princeton, NJ 08544, USA\\
\vskip 0.1cm
$^{c}$Department of Physics, The Ohio State University,\\
 Columbus, OH 43210, USA
     }

\end{center}

\vspace{0.2cm}

\begin{abstract}

We analyze in detail the relation between an exactly marginal
deformation of ${\cal N}=4$ SYM -- the Leigh-Strassler or
``$\beta$-deformation'' -- and
its string theory dual (recently constructed in hep-th/0502086)
by comparing energies of semiclassical strings to anomalous dimensions
of gauge-theory operators in the two-scalar sector. 
We stress the existence of
integrable structures on the two sides of the duality. In particular, we
argue that the
integrability of strings in $AdS_5\times S^5$ implies the
integrability of the deformed
  world sheet theory with real deformation parameter.
 We compare the fast string limit of the worldsheet
action   in the sector with two angular momenta with the
continuum limit of the coherent
state action  of an anisotropic XXZ spin chain describing the one-loop
anomalous dimensions of the
corresponding operators and find a remarkable agreement for all values
of the
deformation parameter. We discuss some of the properties of the
Bethe Ansatz for this spin chain, solve the Bethe equations for
small number of excitations and comment on  higher loop
properties of the dilatation operator. With the goal of going
beyond the leading order in the 't~Hooft expansion 
we derive the
analog of the
 Bethe equations on the string-theory side, and show that they
coincide with the
thermodynamic limit of the Bethe equations for 
the spin chain. We
also compute  the $1/J$ corrections to the anomalous dimensions of
operators with large $R$-charge (corresponding to strings with angular momentum
$J$)  and match them to  the
 1-loop corrections  to  the fast string  energies.
Our results suggest that the impressive agreement between the gauge
theory and semiclassical strings in $AdS_5\times S^5$ is part of a
larger picture underlying the gauge/gravity duality.

\end{abstract}

\end{titlepage}

\setcounter{footnote}{0}

\section{Introduction}


A relation between large $N$ gauge theory and string theory has been
a subject of
active investigation for more than three decades.
The advent of the AdS/CFT
correspondence \cite{MALD,GKPO,WITT}
provided  us  with new  concepts and new tools
to try to put   this remarkable relation on a firm ground.
Still, understanding the  \adss  -- $\N =4$  SYM duality
in detail  beyond the BPS and near
BPS \ci{BMN}  limit remains a challenge.

 It was suggested   in
\cite{GKP2,FT1}  that  concentrating on semiclassical states
with large quantum number (e.g., spin)
 may allow
 one  to verify  the
 agreement between the   large spin  dependence of the
 energies $E$ of strings in
$AdS_5$  and of the anomalous dimensions $\Delta$
of the corresponding SYM
operators.\foot{See  \ci{KRU} for recent progress in this direction.}
 Moreover,  it was  proposed in    \ci{FT} that there  should
exist
a large class of multi-spin
semiclassical string states  and dual
``long'' SYM operators  for which
the   leading coefficient functions
in the large spin expansion of $E$ and $\Delta$ may
  match exactly. This
was indeed confirmed  explicitly, first on particular  examples
 \ci{BMSZ,BFST,AFRT,ART,AS,SEST}    and  then in general
\ci{KRUC,KRTS,KMMZ,KZ}
(see also  \ci{TSE} for a review and additional references).

The underlying reason for  this agreement
(which
 may look unexpected
 in  view of the  different  limits
taken on the string  and the gauge theory  sides)
 appears to be  the equivalence
of the  two
integrable systems  that govern the corresponding
leading-order corrections and
 whose structure happens  to be tightly constrained.
A  conceptually simple  and universal  way
 of understanding   this is by showing the
 equivalence of   (i)
 the  ``fast-string'' limit
of the string action  producing a non-relativistic sigma model action
for ``slow''  string degrees  of freedom, and (ii)
 the ``Landau-Lifshitz''  action
for coherent spin chain  states  describing  the relevant sector
(semiclassical spin wave states with energies  scaling as $1/ J$
in the large spin $J$ limit)  of the ferromagnetic
spin chain  appearing on the gauge theory side \ci{KRUC,KRTS}.

Both the world sheet theory  of strings in $AdS_5\times S^5$ 
and the dilatation operator of the planar ${\cal N}=4$ theory
are described by integrable systems, the latter being an
integrable spin chain \ci{MZ,BKS}.\foot{This spin chain is
explicitly known at the 1-loop level in the full theory \ci{BEST},
to the first few loop orders in various subsectors closed under
renormalization group flow \ci{Be}, and the asymptotic Bethe
ansatz for the $su(2)$ subsector were  proposed in \ci{BDS}
(see \ci{B} for a  review and references). For dilatation operators
and associated integrable spin chains in less supersymmetric
gauge theories see also \ci{D} and references therein. } One  can
utilize the integrability to show that the   integral  equation
and the associated spectral curve
 that classifies  classical solutions of the
string sigma model
(expanded to first two  orders in the ``fast-string''
parameter $\lambda \over J^2$)
is equivalent to the  ``thermodynamic'' (large $J$)
limit of the  one- and two-loop  Bethe ansatz  equations
appearing on the gauge theory side  \ci{KMMZ,KZ}.
This relation between the classical ``string Bethe equation''
and the thermodynamic limit of the spin chain Bethe ansatz
(which  appears to extend  also to  the first subleading $1/J$
level  \ci{BTZ})
provides an alternative demonstration of the
 ``$E=\Delta$'' equivalence. It
 also  prompts to try to find an analog  of an interpolating Bethe
ansatz that would describe the spectrum
of free strings in \adss or of  the  corresponding SYM operators at any
finite `t Hooft coupling $\l$ \ci{AFST,Staud}.

\bigskip


Given this  remarkable  progress in understanding
\adss string -- $\N=4$ SYM  duality  one would  like
to try to extend it to  more ``realistic'', i.e.  less supersymmetric,
examples, anticipating
that the  semiclassical
string  states  should  be  capturing some
general  features of the correspondence between planar gauge theory
and free  string theory.
 It is natural  to   start  with
examples   where  the gauge theory side is superconformal
at {\it  any}  value of  the `t Hooft coupling $\l$. In such a case
one  would have, as  in $\N=4$ SYM,  a luxury  of
being able to use the gauge theory perturbative expansion
at  small $\l$ and the string theory perturbative expansion
at large $\l$ and then to compare the results  for semiclassical  states
with large $J$ in an appropriate scaling limit.
 One obvious class of such examples
are orbifolds of the \adss string  dual to orbifolds of
  $\N=4$ SYM  \ci{KS} (see \cite{ide})
 but they are somewhat trivial  not involving a  new continuous
parameter:   we would like to have  an adjustable
{\it continuous} deformation parameter  which could  allow us
to define new scaling limits.

A non-trivial starting point would be the  well-known
one-parameter   family of $\N=1$ gauge theories
which are  exactly
marginal   deformations \ci{LEST}  of the \sym   theory
(for earlier work on $\N=1$ superconformal theories see \ci{old}),
but until very recently \ci{LUMA} the corresponding dual type IIB
supergravity  background  and thus the  dual
superstring theory  was not known explicitly.\foot{A perturbative approach
to finding  the corresponding supergravity background was developed in
\cite{AKYA}.}
 The remarkable observation made
in \ci{LUMA} is that
 in the case of a real deformation  parameter $\b$
 the dual 1/4 supersymmetric supergravity background  can be
constructed  by applying a combination of T-duality, shift of angle
and another T-duality (``TsT''-transformation)
to the original  \adss  background.\foot{Similar transformations were
 used in  \ci{RT} to construct Melvin-type magnetic flux tube
backgrounds {}from flat space and also to solve the
corresponding  string theory.}
The resulting  string theory in this background,
whose metric  is a product of $AdS_5$  and a
5-space with topology of $S^5$ and which depends, in addition to the radius
$R= \sqrt{\a' \sqrt \lambda}$,  also on the  new deformation  parameter
$\b$,
may then be viewed as an exactly marginal deformation (now
in the 2d sense)  of the \adss superstring theory.
One may  also try to  generalize to   the case of the
non-zero   imaginary part of
the parameter $\b$ by applying
an extra $S$-duality transformation to the TsT transformed
\adss  supergravity  background. However,  in this case
the supergravity background  may   receive
non-trivial  $\a'$-corrections
and thus the corresponding string theory  may not be possible
to define in a useful  closed form (see also below).

\bigskip

Our aim  in this paper  will be to study the duality example
of \ci{LUMA} by  comparing semiclassical
string states to a class of  gauge theory scalar operators
whose 1-loop anomalous dimensions are again described
by an integrable spin chain \ci{RR,BECH}.
Surprisingly, we will find that many of the results that
were obtained  in  the undeformed $\b=0$ case of \adss -- \sym
duality have a straightforward  generalization  to the
case of (real) non-zero $\beta$.
 There is again  an equivalence
between  the   ``fast-string'' limit of the classical string  action
and the  Landau-Lifshitz action associated to the 1-loop spin chain,
i.e. the matching between leading order terms in the string energies
and anomalous dimensions of the dual gauge theory operators.
There is also  an analog of  classical ``string Bethe equations''
of \ci{KMMZ}.

\bigskip
\bigskip


We shall begin  in \S\ref{rev}
by recalling the form   of the exactly
marginal
deformations of  the ${\cal N}=4$  SYM \ci{LEST}
and their gravity duals \cite{LUMA}.
We shall
emphasize the string theory consequences of the T-duality
algorithm used to construct them.

To be able to  compare  string theory and
gauge theory results it
is important to understand the relation between the parameters
appearing in the two theories and their scaling properties
in the sector of states
with    large quantum numbers. We will discuss this in
\S\ref{limpar}.

One of the consequences of the  TsT transformation
used to find the  string theory dual of the  deformed gauge theory
is that  for each semiclassical string solution
in $AdS_5\times S^5$ there exists the  corresponding
one in the deformed background.\foot{For integer shift parameter
the TsT transformation  is a symmetry of the
\adss string theory,
i.e. while the two supergravity backgrounds still
look very different,
the corresponding string spectra must be  isomorphic (cf. \ci{TTT}).
This is  true also in the case when one performs
 shifts of the 3 isometric angles of $S^5$ by  2 integer  parameters.
}
 In
 \S\ref{LL_action}  we shall first  present  an example of such
a solution
and then  follow  \ci{KRTS,KT}  to
find
an  effective action describing all such solutions in the
large angular momentum  limit.
The  corresponding equation of motion expanded  to
the leading order in  large  angular
momentum  is a ``twisted'' version
of the (anisotropic for complex $\b$)
 Landau-Lifshitz equation
which  is  known to be integrable.

Next, we shall   turn  in  \S\ref{coh_state_action} to
 the $\N=1$ supersymmetric
gauge theory side,   starting  with  the spin chain
description of the 1-loop dilatation operator in the deformed
gauge theory
in a scalar field  sector
\cite{RR,BECH}.
We shall then  analyze the
continuum limit  of the coherent state expectation value  of the
integrable   spin chain Hamiltonian
and construct the corresponding Landau-Lifshitz
effective action  for low wave-length spin  excitations.
As in the undeformed case, this
 action happens to be exactly  the same
 as found on the string side. This implies matching  between
string energies and 1-loop gauge theory anomalous dimensions
and also matching of the  associated integrable structures.

 In the following section
we shall  present the Bethe ansatz   for the
integrable spin chain representing the 1-loop  dilatation operator
of the deformed theory.
We shall then
 explicitly solve the
 Bethe equations in the small excitation
(or near-BPS or ``small string'')
limit,
i.e. in the  regime corresponding to a particular  plane
wave limit of the geometry \cite{LUMA}.
We shall also discuss  higher-loop corrections  and point out that
while the 1-loop dilatation
operator in the  2-scalar sector is integrable for generic
deformation, the  2-loop correction  preserves integrability
 property only for real
deformations.

In section 7, with a motivation
to  go beyond the leading order,
 we will  show how the derivation
of the string Bethe equation describing classical solutions
of strings moving on $S^3$ \ci{KMMZ}  generalizes  to the case of real
$\b$ deformation.
We shall use the Lax representation  constructed in \cite{SF}
which implies  integrability of
strings moving in  the background of \ci{LUMA}.
This  integrability property   is remarkable  since the target space
of the deformed sigma model is  no longer a coset space.

In \S 8 we shall  compute, following the discussion of
$\b=0$ case in \ci{BTZ},  the 1-loop subleading $1/J$ corrections
in the 2-spin sector  and
demonstrate that
the correspondence between the gauge and string theory
results holds also  in the deformed theory.

Our conclusions will be summarized in \S\ref{con}.

In Appendix A we shall discuss the relation between
the spin chain Hamiltonian  used in this paper and a  more general one
considered in \ci{RR}.
In Appendix B
we shall give some details on asymptotics of
 the monodromy matrix used in section 7.

\renewcommand{\theequation}{2.\arabic{equation}}
 \setcounter{equation}{0}

\section{Deformed ${\cal N}=4$ SYM  and its string theory dual
\label{rev}}

The ${\cal N}=4$ super-Yang-Mills theory exhibits two exactly
marginal deformations \cite{LEST}.
The corresponding terms in the superpotential
  in the ${\cal N}=1$ superspace  action
are $\Tr[\Phi_1\{\Phi_2,\,\Phi_3\}]$ and $\sum^3_{i=1}\Tr[\Phi_i^3]$.
In the following we will be interested only in the first type of
deformation.

\subsection{$\beta$-deformed ${\cal N}=4$ SYM theory: a brief review
\label{def_action}}

In the notation of \cite{LUMA},
the superpotential of ${\cal N}=4$ SYM theory deformed with
operators of the first type is
\begin{equation}
\label{supe}
W= \, h\ \Tr( e^{ i  \pi \beta} \Phi_1\Phi_2 \Phi_3 - e^{- i  \pi \beta}
 \Phi_1\Phi_3 \Phi_2  )\ , 
\end{equation}
where $\beta$ is allowed to have an imaginary part. We will call this
theory {\it  $\b$-deformed ${\cal N}=4$ SYM}.
Our 
notation for the deformation  and coupling parameters is 
  related to the one in  \cite{LUMA}
as follows:
\bea
\beta&\equiv &\beta_d-i\kappa_d\ ,
~~~~~~~~
\beta_d=\gamma-\tau_{1s}\sigma_d\ ,
~~~~~~~~
\kappa_d=\tau_{2s}\sigma_d\ ,\la{kkk}\\
\tau_s&=&\tau_{1s}+i\tau_{2s}=\frac{\vartheta}{2\pi}+\frac{4\pi i}{g^2_{\YM}}\ .
\label{not0}
\eea
If $\b$ is real, i.e. $\s_d=0$ we  have
\be \beta_{_{\sigma_d=0}}=\gamma  \ . \ee
We   use the subscript $d$ to indicate ``deformation''
and the subscript $s$ to indicate ``string coupling'' (in an attempt to
avoid confusion with the worldsheet coordinates $\tau$ and $\sigma$).

The deformation leading to (\ref{supe}) preserves the Cartan
subalgebra of the $SO(6)$  $R$-symmetry of ${\cal N}=4$ SYM and the diagonal
$U(1)$ is the $R$-symmetry of the manifest ${\cal N}=1$ supersymmetry
algebra. In its general form (\ref{supe}) does not lead to a
conformal field theory; 
requiring  conformal invariance
gives a constraint relating $h$,
$\beta$ and the gauge coupling constant.
To two-loop order this constraint can be found using the
general  results about finiteness of $\N=1$ supersymmetric gauge
theories \ci{old}.
In the large $N$ limit it reads
(we assume  normalizations such that $h=1$ in $\cal N$=4
SYM theory)\foot{We are grateful to D. Freedman  for
 pointing out an
error  in this condition in the original version of this paper.}
\be \la{confcond}
|h|^2 \  ( |e^{i\pi \beta}|^2 + |e^{-i\pi \beta}|^2)  =2\ .
\ee
For real $\beta$   the large $N$  conformal invariance condition
$|h|^2 =1$ turns out to be exact to all loop orders \ci{za}.

Integrating out both
the $F$ and $D$ auxiliary fields and restricting  to the scalar
sector, one finds the following   
scalar potential\foot{We use the same notation
$\Phi_i$ for the first component of scalar superfields in
\rf{supe}.}
\begin{eqnarray}
V= 
&& |h ^2{\bf q}^{-1}|\  \Tr\Big[~|\Phi_1\Phi_2
-{\bf q}\Phi_2\Phi_1|^2+|\Phi_2\Phi_3 -{\bf q}\Phi_3\Phi_2|^2
+ |\Phi_3\Phi_1 -{\bf q}\Phi_1\Phi_3|^2 \Big]\vphantom{\Big|}\nonumber\\
&&~~~~~~~~+\  \Tr\Big[~([\Phi_1,\,{\bar \Phi}^1]+[\Phi_2,\,{\bar
\Phi}^2]+[\Phi_3,\,{\bar \Phi}^3])^2\Big]\la{vv} \\
{\bf q}&\equiv& e^{-2\pi i \beta}\ , ~~~ 
~~~q_d\equiv |{\bf q}| =e^{-2\pi\kappa_d} \ .
\label{vev}
\end{eqnarray}
When  ${\bf q}$ is a primitive $n$-th root of unity  it  was
known for a while \cite{BELE} that the dual geometry is that of the
near-horizon D3-branes on orbifolds with discrete torsion. Until the
recent work of \ci{LUMA}
 it was  not clear, however,  what  is the gravity dual  for the
general {\it continuous} deformation parameter.

Below we  will not review the derivation and results of \cite{LUMA} in detail
concentrating instead
on the world sheet implications of their construction.

\subsection{Deformed type IIB supergravity
background  \\ and  the  corresponding
  superstring theory
\label{deformed_bkg}}

The supergravity background dual  to the
$\beta$-deformed SYM theory   was constructed in \ci{LUMA}
using a
combination of T-dualities  and a  shift
on
the isometries of the 5-sphere part of  $AdS_5\times S^5$.
In  addition, a  non-trivial  $S$-duality
transformation was  necessary   for the construction of the
background dual to the gauge theory with {\it  complex}  deformation
parameter $\beta$, i.e. with $\s_d\not=0$.

Let us begin with the simplest case of
 the real deformation $\b$ (i.e. when $\bq$=pure phase, \ $q_d=1$).
 Since  in this case the  construction of the supergravity background
requires only the use of T-duality
and a shift,
we can
implement it  directly at the level of
 the world sheet string theory.
The starting point should then be
the classical Green-Schwarz action in $AdS_5\times
S^5$ \cite{GSclass}.\foot{For the purpose of eventually considering a  quantum string  theory
the  starting point may be  the light-cone $\kappa$-symmetry
gauge fixed
action of \cite{MTTH}; this
action  also has a  desirable feature that
the isometries necessary to perform  the T-duality transformations
\cite{LUMA} are not obscured by the gauge fixing condition.}

Through the usual implementation of T-duality as a 2-dimensional
duality (with the well-known  quantum dilaton shift
properly taken into account \ci{BU}),  or directly using the
T-duality transformations spelled out
in the context of the Green-Schwarz string in \cite{TDGS,TKR},
it is then not too   hard  to find the
world sheet type IIB  superstring action resulting
{}from applying the TsT
transformation to the \adss  string action.
The existence of such  direct  world-sheet
construction of the ``$\b$-deformed''  superstring action
exposes three  important facts:

(1) Since the  T-duality
 and the coordinate transformations should
preserve the 2-dimensional conformal invariance of the
string theory,
there should exist  a renormalization scheme in which the
space-time
background obtained in this way does not receive $\alpha'$ corrections
even though it preserves only eight \ci{LUMA} supercharges
(3/4 of supersymmetries are  broken if $\beta$ is not  integer).

(2) The 2-dimensional duality maps solutions of the original string
equations of motion into solutions of the dual  equations of
motion;  this construction then  represents  a formal way of
generating solutions of the classical equations of motion
for a string moving in  the dual
string background.
One should still take care of the global issues like   periodicity
 of world sheet coordinates  by appropriate twists
which implies  that the original
and the TsT transformed string theories
are not actually equivalent
for non-integer $\beta$.

(3)
The construction of the one (spectral) parameter  family of
flat currents or the  Lax pair  which
demonstrates   the classical
integrability of the free
string theory  in $AdS_5\times S^5$ relies only  on
the classical string equations of motion \ci{MSW,BPR,VALL,POLY}.
This implies
 that at least part  (in fact, an infinite subset) of the
integrals of motion of \adss sigma model
 are mapped into integrals of motion  of the deformed sigma model,
 suggesting that  in the case of real $\b$ the
deformed string theory is also classically integrable.\foot{After all,
a non-technical meaning of integrability
of a (2-d)  field theory  is that there exists at least an implicit but systematic
way of
classifying  all classical solutions.
The usual
  necessary condition for that
 in a field-theory context  where there is an infinite number of degrees of freedom
is the
existence of an infinite number of conserved charges. This
practical criterion
is certainly satisfied  for this    deformed string theory.}

The resulting
supergravity background for  string theory dual of  the
real $\b$-deformation  of the ${\cal N}=4$ SYM is \ci{LUMA}:
\begin{eqnarray}
ds^2_{\rm str} &=& R^2  \left[ ds^2_{_{AdS_5}} +
   \sum^3_{i=1} ( d\rho_i^2  + G \rho_i^2 d\phi_i^2) +   \tg ^2G
\rho_1^2\rho_2^2\rho_3^2 (\sum^3_{i=1} d\phi_i)^2 \right]
\label{metrgen0}
\\
\la{bfield0}
B_2 &=&  R^2\tg  G w_2 \ , \ \ \ \ \ \ \
 w_2  \equiv   \rho_1^2 \rho_2^2 d\phi_1 d\phi_2 +
\rho_2^2 \rho_3^2 d\phi_2 d\phi_3 + \rho_3^2 \rho_1^2 d\phi_3 d\phi_1\ ,   \\
e^\phi &=& e^{\phi_0}G^{1/2} \, ,\ \ \ \ \ \    \qquad\chi = 0 \, ,
\\
\label{ofggen0}
G^{-1} &\equiv& 1 + \tg ^2   Q  ~,\qquad
\ \ \ \ \ \
Q \equiv  \rho_1^2 \rho_2^2 + \rho_2^2 \rho_3^2 + \rho_1^2 \rho_3^2
~,\qquad \sum_{i=1}^3\rho_i^2=1 \ ,
\\
 C_2  &=& -  12R^2 e^{-\phi_0}  \tg  w_1 d\psi
\ , \ \ \ \ \ \ \ \ \ \ \     d w_1 \equiv     \cos \alpha
\sin^3\alpha \sin {\theta}  \cos {\theta}
d\alpha d\theta ~,
\\
F_5 &=& 4 R^4 e^{-\phi_0}  ( \omega_{_{AdS_5}} +  G \omega_{_{S^5}})
~,\qquad\omega_{_{S^5}} \equiv { }  dw_1 d\phi_1 d\phi_2 d\phi_3
\end{eqnarray}
Here $B_2$ is the NS-NS 2-form potential, $\phi$ is the dilaton
and $d\chi, \ dC_2 $ and $F_5$ are the R-R field strengths.
 The angles $\psi$ and  $\theta$,$\alpha$  appearing in $dw_1$
are defined as follows
\be\la{angl}
\psi= {1 \ov 3} (\phi_1 + \phi_2 + \phi_3) \ , \ \ \ \ \ \ \
\rho_1 = \sin \a \cos \th\ , \ \ \
\r_2 =  \sin \a \sin \th\ , \ \ \
 \r_3=  \cos \a\
 .
\ee
The standard \adss background is recovered  after setting the deformation
parameter $\tg = R^2 \g \sim \b$  to zero (then $G=1$, $\p=\p_0$
 and $B_2=C_2=0$).
 The parameter $\tg$ was  denoted as $\hat \g$ in \ci{LUMA} and
other parameters are
\bea
g_s &=&  e^{\phi_0} = {1\ov \tau_{2s}} =
{g^2_{\YM}\ov 4\pi}\, , \la{o} \\
\quad R^4&=&e^{\phi_0}R_E^4=4 \pi g_s N =N
g^2_{\YM} \equiv\lambda \ , \ \ \ \ \ \ \ \ \   \alpha'=1\ , \la{p}   \\
\tg &=&   \g R^2\  .\la{fg}   \eea
In order for the  superstring theory  defined on this background
  to have a consistent
perturbative  expansion,  the
 parameter $\tg$ (cf.\rf{not0}) and thus the string-theory action  should
 be independent of the string coupling constant $e^{\p_0}$.
Indeed,
observing that the superstring action
depends on the RR field strengths  $F_3=dC_2$ and $F_5$  in combination
with
$e^{\phi}$ factor  \cite{TSDD}
and also
that the vielbein components of $F_3$ and $F_5$ both scale as $R$
we conclude that the inverse string tension
expansion
of the classical ($g_s=0$) superstring action
in the above background will depend only
on
 $R^2=\sqrt \lambda $ and the parameter $\tg$ but not on  $
e^{-\phi_0} = { \tau_{2s}}$.
This is  what is needed
to be able to use a systematic semiclassical approximation
to string theory, with $\tg$ playing  the  role of a regular
deformation parameter of the  supergravity background and of   the
 superstring  action.

\bigskip

Next, let us follow \cite{LUMA} and apply S-duality\foot{The required
$S$-duality transformation depends on one real parameter, $\sigma$, 
and acts
 on the complex coupling constant
 $\tau = \chi_0 + i e^{-\phi_0}$, thus
introducing the two new real
parameters $\chi_0$  and $\sigma$.}                                                                                              
before and after the
TsT-transformation leading to the above type
IIB supergravity 
background.\foot{It is important to note that
 since 
there are two S-duality transformations, the initial and final
backgrounds are at weak string coupling.}
 This step may be viewed as a formal trick
 to generate a type IIB supergravity  solution
 dual to 2-parameter (complex $\b$)
 deformation of the $\N=4$
SYM theory.\footnote{It may still  be of interest to study
the  interplay between the corresponding fundamental string and D-string
actions that probe, respectively, the S-dual backgrounds. In general,
 starting with a D-string
in a given background and ``rotating away'' the 2d gauge field  by a 2d
duality transformation
one  ends up with the fundamental string action coupled to the  S-dual
background  \cite{TSDS}.}
We finish with the following expressions (equivalent to the ones in
\cite{LUMA}
but written for the string-frame metric and using
 the notation introduced
above):
\begin{eqnarray}
ds^2_{\rm str} &=& R^2 H^{1/2} \left[ ds^2_{_{AdS_5}} +
   \sum_{i=1}^3 ( d\rho_i^2  + G \rho_i^2 d\phi_i^2) + (\tg^2 + \ts^2) G
\rho_1^2\rho_2^2\rho_3^2 (\sum_{i=1}^3 d\phi_i)^2 \right] \label{metrgen}
\\ \la{bfield}
B_2   &=&  R^2\left(\tg  G  w_2  -
    12\ts w_1 d\psi\right)\ ,
\\
e^\phi &=& e^{\phi_0}G^{1/2}H \, ,\qquad\chi = \chi_0+e^{-\phi_0}
\ts\tg H^{-1}Q \, ,
\\
\label{defofggen} G^{-1} &\equiv& 1 + (\tg^2 + \ts^2)   Q  ~,\qquad
H   \equiv   1 + \ts^2 Q \ ,
\\ \label{gzeroe}
 C_2  &=& R^2\big[ ( \chi_0  \tg -  e^{-\phi_0}  \ts )  G w_2
-  12( e^{-\phi_0}   \tg + \chi_0  \ts )w_1 d\psi\big] \ ,
\\
F_5 &=& 4 R^4 e^{-\phi_0} ( \omega_{_{AdS_5}} +  G \omega_{_{S^5}})
~.
\end{eqnarray}
Here $\ts$ is the new deformation parameter and we have  used the
following definitions
(cf.  \cite{LUMA} and \rf{not0}):
\bea
\chi_0  + i e^{-\phi_0} &=& \tau_s = \tau_{1s} + i \tau_{2s}
= \frac{\vartheta}{2\pi}+\frac{4\pi i}{g^2_{\YM}}\ , \\
\tb &=& \tg - i\ts = \beta R^2=(\g - \tau_s\s_d)R^2\, ,\\
\tg &=& (\g - \tau_{1s}\s_d)R^2=\beta_d R^2 \, ,\ \ \ \ \ \ \  \ts =
\tau_{2s}\s_d  R^2=  \kappa_d R^2\ .
\label{ident3}
\eea
We shall assume that  the parameter $\tb$ is
independent of the string
coupling constant.

The classical  superstring action in  this  S-dual  background
is  straightforward to write down, at least to quadratic order in
fermions.\foot{Note that
here the string metric  is no longer a direct product of $AdS_5$ and $S^5_\b$
but rather  conformal to it. That means that in general the
classical string motions in $AdS_5$ and $S^5_\b$ parts  will no longer factorize.}
Here, however,    we have no good reason to believe  that the S-dual
background invariant
 only  under  eight supercharges
 will  not be deformed by $\a'/R^2$
corrections:
 we may then need to modify the classical
superstring action   by extra $\a'/R^2$
correction terms
 in order to
ensure its quantum 2d conformal invariance.
These extra terms  may depend also on background
 field strengths and thus on other parameters like $\p_0$ and $\chi_0$.
For simplicity,  let us  assume
that this does not happen.
Then the ``NS-NS'' part of the  superstring action
will obviously depend only on  $R^2$ and
$\tb$ (more precisely, on $\tg$ and $\ts$), i.e. will have
no dependence on the string coupling constant $\tau_s$.
As in the $\s_d=0$ case discussed above, this
 is important  in order
 to be able to develop  a semiclassical
expansion
for the energies of the corresponding string states
which we  will  be interested in.

One may  still worry (legitimately) about the    $
e^{-\phi_0} $  and $\chi_0 $  dependence of  the
fermionic couplings to the  RR field strengths. The fundamental
string action is
certainly
not invariant under the  S-duality applied to a background
in which  it is moving.
However,   to quadratic order in fermions
(which is enough to study 1-loop correction to semiclassical
approximation)
one is able to argue  that
this dependence is fake and just reflects the
 construction via S-duality transformation:
the superstring fermions couple
though the generalized covariant derivative
to a particular combinations of fluxes  which is  covariant under the
S-duality.
Indeed, this generalized covariant derivative  is the same
as the one appearing in the type IIB supergravity  variation
of the
gravitino field,  and the latter   (like the full type IIB
supergravity action)
is invariant under S-duality.
 Explicitly,
the RR 3-form field strength $dC_2$  enters  the quadratic fermionic action
 though the combination
$e^\phi \left(dC_2 - \chi  d B_2\right)$, and
computing this combination  on the above background we
observe   that the dependence  on the dilaton
and the RR scalar  constants $e^{\phi_0}$ and $\chi_0$
indeed cancels out. The spinor covariant derivative
contains also  the RR scalar field strength
  term $e^\phi d\chi$,   and it  is again independent of
$\phi_0$ and $\chi_0$.

We conclude that, at least  {\it to  quadratic order in
fermions}, the corresponding superstring action
does depend only on the overall factor of string tension
$R^2$ and on the parameters $\tg$  and $\ts$ but not on the string
coupling $\tau_{s}$. The general case  of the full string action is,
however, unclear, i.e.  the perturbative semiclassical study
of the gauge-string duality   in  case of complex $\b$ deformation
appears to be problematic.


\renewcommand{\theequation}{3.\arabic{equation}}
 \setcounter{equation}{0}

\section{Semiclassical string expansion:
limits and parameters \label{limpar}}

In the semiclassical approach to string spectrum
one starts with a classical string  action on a 2d cylinder
and computes the energies  of classical string solutions.
One may then compute also
 quantum ($\a'/R^2$) string sigma model corrections,
for which one will need to include  also the fermionic
  terms in the string action (and, in general,
 to take into account
the dilaton coupling, cf.\ci{papa}). In the  case of multispin  string
states considered  in \ci{FT} the quantum string corrections were
suppressed in the large spin limit ($J \to \infty$), and the same  is
expected to happen in the present ``deformed'' case.

The relevant
bosonic part of the classical  string action
 depends on the metric $G_{MN}$ and the NS-NS 2-form
field $B_{MN}$, i.e. is given by
\bea
\la{A}
S_B = -{R^2\over 2}\int\, d\tau\int  {d\s\over
2\pi} \left[ \g^{\a\b}\partial_\a X^M\partial_\b X^N\, G_{MN}
-
\e^{\a\b}\partial_\a X^M\partial_\b X^N\, B_{MN}\right]\, ,
\eea
where we set $\alpha'=1$, $\e^{01}=1$,
and $\g^{\a\b}\equiv \sqrt{-h}\, h^{\a\b}$\
($h^{\a\b}$ is a world-sheet metric with Minkowski signature, i.e. in
the conformal gauge $\g^{\a\b} = \mbox{diag}(-1,1)$).
The string-frame metric
 and the $B_2$-field are given by
(\ref{metrgen0}) and (\ref{bfield0}) or, for complex $\beta$, by
(\ref{metrgen}) and (\ref{bfield}).

The parameters  in
the string sigma model (\ref{A}) are then
the $AdS_5$ radius $R^2 = \sqrt{\l}$
(playing the role of
the coupling constant of the sigma model)
 and the complex deformation
parameter $\tb$ or, equivalently,
 the two real parameters  $\tg$ and $\ts$.
Note that since the metric and $B_2$-field
 have   regular
expansion in powers of  $\td \g$  and  $\ts$,
the same is true for the classical string action,
with the zero-order  term
being the \adss bosonic sigma model   with the $S^5$ part being
  $\int \sum^3_{i=1}
[ (\del\rho_i)^2
+ \rho^2_i (\del \phi_i)^2 ] $.

The semiclassical approximation  \cite{GKP2,FT1} is useful for  states that
carry large energy and which can thus be approximated by
classical string solutions  with $E= \sql \E + ... \gg 1$.
Here $\E$ (depending on parameters of  classical solution)
is fixed in the semiclassical expansion,
and quantum corrections  are small provided
the sigma model coupling constant  $\sqrt{\l}$ is large.
In present AdS/CFT context, the energy  of the corresponding
 quantum string states
will in general involve functions of $\l$ interpolating between
the  ``perturbative string theory''
($E = \sql (\E + {b_1 \ov \sql} + ...) $)
and the ``perturbative gauge theory'' ($E = c_1 \l + c_2 \l^2  + ... $)
regimes.
However, for a  special (large) class of
semiclassical \adss  string states such interpolation functions
happen to be absent in the first few leading terms
of the expansion of the energy in large quantum numbers \ci{FT}.

Here we would like to study  the  analog of this class of multi-spin
states in the $\b$-deformed theory. The string solutions
of the type we are going to consider below\foot{More generally,
the relevant string  configurations  are ``fast strings'' \ci{MIHA}
that may not only rotate  but also pulsate \ci{MINA}, etc.}
have large
total angular momentum $J = \sqrt{\l}{\cal J}$ in (deformed) $S^5$ directions
and their
 energy  admits a regular expansion  in powers of
the effective coupling constant
\be {\tilde\lambda}=\frac{\lambda}{J^2}={1\ov \J^2} \  \ee
which,  in the semiclassical expansion,   is assumed to be
fixed and independent of $\l$.
Since the field theory perturbative  expansion for  anomalous
dimensions of the corresponding  operators
 happens to have the same dependence on both $\l$
{\it and }  $J$,
one may then attempt to  compare the string and field theory expressions.

The string solutions in  the $\beta$-deformed background
will depend  on the deformation
parameters $\tg$ and  $\ts$.
 Since the deformation is smooth,  the classical
energy of a multi-spin  solution
should then have the following expansion\footnote{
The string energy is  real and
 depends  separately on $\tg$ and $\ts$,
and the  dependence on each of the two parameters
is regular since the background fields in (\ref{metrgen}),(\ref{bfield})
admit  regular expansions in $\tg,\ts$.
To simplify the
expressions, instead of writing $E$  as a series in $\tg$ and $\ts$
 we shall   write it  symbolically as a series in
 $\tb$.}
\begin{eqnarray}
E\= \sqrt{\l}\ \E({\tilde\b},{\cal J})=\sqrt{\l}\left[\E _0 ({\cal
J})+ {\tilde\b}\,f_1({\cal J})+{\tilde\b}{}^2 \, f_2({\cal
J})+{\tilde\b}{}^3 \, f_3({\cal J})+\dots\right]\, ,
\label{gen_class_energy}
\end{eqnarray}
where $\E_0({\cal J})$ represents the energy   in the
$AdS_5\times S^5$ case, i.e. it has   regular
 large $\J$ or small $\tl$
expansion  \ci{FT}
\bea\la{clas_energy_adss}
E_0 \= \sqrt{\l}\E_0({\cal J}) =J f_0(\tl)=
J\left( 1 + c_1\tl +c_2\tl^2 +\dots\right)\, .
\eea
To be able to compare to gauge-theory perturbative expansion, we
would like  to find out
 when the $\tb$-dependent terms in (\ref{gen_class_energy}) admit
a similar expansion.
To this end we notice that $\tb\equiv R^2 \b $ in \rf{ident3}
can be written in the form
\begin{equation}
\la{tb}
{\tilde\b}
=\frac{R^2}{J} \ \b J   = \frac{\sqrt{\l}}{J}\ \b J = \sqrt{\tl}\ \b J~,
\end{equation}
where $\b$ is the deformation parameter on the gauge theory side.
Being a parameter of
the classical string action
 $\tb$ is  fixed in  the semiclassical expansion
in which $\tl$ is also  fixed;
it then  follows that  $\b J $  should also  be kept fixed
in the limit $J \to \infty$.
Equivalently, we need to assume that the string-theory perturbative  expansion
corresponds to the limit when $J \to \nfty$, \ $\l = \tl  J^2 \to \infty$ and
$\b = {\tb \ov \sqrt{\tl} J} \to 0$.

We   will take into account that   $\b \sim { 1 \ov J}$
corresponds to the  ``semiclassical string''  scaling
limit in the subsequent sections
 when we will (i) define  the  relevant
 continuum  limit of the spin chain Hamiltonian  which represents
 the one-loop
dilatation operator in the ``$su(2)_\b$'' subsector of the
 $\b$-deformed gauge  theory, and  (ii)  when  we will consider the
thermodynamic limit of the Bethe ansatz for
the spin chain. This  will amount to isolating  a class
of ``macroscopic'' spin-wave   states or gauge theory operators
which are dual to  the semiclassical string states.

Using the relation (\ref{tb}), we can rewrite the
expression for the classical energy
(\ref{gen_class_energy}) as follows
\bea
\la{clas_en}
E\= J\left[f_0(\tl)+ \tl \b J \, f_1({\cal J})+\tl (\b J)^2 \,{f_2({\cal J})\ov \J}
+\tl^2 (\b J)^3 \, f_3({\cal J})+\dots\right]\, .
\end{eqnarray}
In general, the coefficient functions $f_1,f_2,...$
may have  an expansion in powers of
$\sqrt{\tilde\lambda}$,
but in order   for a semiclassical string energy
to have a perturbative gauge theory interpretation it is necessary
that \rf{clas_en}
 has a regular expansion in powers of ${\tilde\lambda}$.
If this is the case,
 $E$  may  be formally rewritten
 as a  series of terms containing powers
of $\l,\b$ and $1\ov J$, i.e. in the same form
which one expects to find for a perturbative  anomalous dimension
on the gauge-theory side.
This boils down to the condition that
$f_1,\J^{-1}f_2,f_3,...$  should have regular expansions in powers of
$1/\J^2= \tl$. Assuming this one finds
\begin{eqnarray}
E=J\Big[
1 + {\tilde\lambda}\left( c_1 \+ d_1\, \b J\+ d_2\, (\b J)^2 \right)+
O(\tl^2)
\Big]\ . \la{en}
\end{eqnarray}
As a consequence, we observe that comparing to the  gauge-theory
1-loop  anomalous dimensions we will  be able to  probe
only  the terms of order $(\b J)^k$ with $k\le 2$.

To  be able to compare  the terms of higher orders  in $\b J$
 one should take into account the world-sheet quantum corrections.
Computing quantum string  corrections does not pose a problem
in the case of real $\b$ or $\s_d=0$, but
this is by far less clear in the general case of $\s_d\not=0$.
As
 we have pointed out  in the previous section,
 in the case  of an arbitrary complex deformation
 the RR fluxes depend explicitly on the string coupling
$\tau_s$, but  this dependence drops out of   the action at the
quadratic fermionic level; then    there should be  no ``anomalous'' string
coupling dependence at least at  the 1-loop string sigma model
level.
Then the semiclassical  loop expansion for the string energy should
look like:
\begin{eqnarray}
E=\sqrt{\l}\E_0({\tilde\b},{\cal J})+\E_1({\tilde\b},{\cal
J})  + \dots\, .
\end{eqnarray}
In particular,  including
the  world sheet one-loop  quantum
correction allows one  to compare
 with  gauge-theory  1-loop   terms
the terms in  $E$  that contain   factors of
$(J\b)^3$ with $k\le 3$.


\renewcommand{\theequation}{4.\arabic{equation}}
 \setcounter{equation}{0}

\section{
Strings on $R_t\times S^3_\b$ and  large spin
limit of the string sigma model
\label{LL_action}}

In this section we shall consider classical
strings located at the center of $AdS_5$ and moving
in the ``$S^3_\b$'' part of the deformed 5-space in \rf{metrgen}
defined by the condition (see \rf{angl})
\be \la{kopp}
\alpha= {\pi\ov 2}  \ , \  \ \ \ {\rm i.e. }  \ \
\ \ \  \
\r_1 = \cos \th \ , \ \ \ \
\r_2= \sin \th \ , \ \ \   \ \rho_3=0\ ,\
\ee
which, as it is easy to see,  is indeed
 a consistent truncation of the string
equations of motion.\foot{More precisely, in the case of $\ts\not=0$
the space where string moves is not a direct product
$R_t\times S^3_\b$  but is conformal to  it.}
The resulting bosonic string action \rf{A}  is
\begin{eqnarray} \la{a1}
 S = - \ha {R^2}\int d\tau
\int {d\s\over 2\pi} \Big[ &&\!\!\!\!\!\!\!
\g^{\a\b} H^{1/2} \left( -\pa_\a t\pa_\b t  +
\partial_\a \th \partial_\b \th
+ G\, \cos^2  \th\ \partial_\a \phi_1\partial_\b \phi_1
+ G\, \sin^2  \th\ \partial_\a \phi_2 \partial_\b \phi_2  \right)   \nonumber \\
&&\!\!\!\!\!\!\!\!\!\!\!\!
-\  2 \e^{\a\b}\hg\, G\,
\sin^2\th \ \cos^2\th\ \partial_\a \phi_1\partial_\b\phi_2 \Big]\ ,
\end{eqnarray}
where
\be\la{gh}
G= \big[1 + \four (\tg^2 + \ts^2) \sin^22\th\ \big]^{-1} \ , \ \ \ \ \ \ \
H= 1 + \four \ts^2    \sin^22\th  \ .
\ee
and we recall that the string theory deformation parameters ${\tilde
\gamma}$ and ${\tilde\sigma}$ are related to the gauge theory
deformation parameters $\gamma$ and $\sigma_d$ by the equations
(\ref{ident3}) and (\ref{not0}).
To get an  idea of
how the $\b$-deformation may influence
the form of string solutions  we shall first
discuss  a particular example  which is
a  deformation of the simplest 2-spin circular string
solution of \cite{FT,ART}.

We shall then consider, following what was done in $\b=0$ case in
\ci{KRTS,KT},   the
 string sigma model on $R_t \times S^3_\b$
in a particular gauge where  $t \sim  \tau$
and the density of the angular momentum
 in the two angular directions $\phi_1$ and $\phi_2$
is constant. This uniform gauge allows  one to develop
a systematic large spin or ``fast-string'' expansion,
with the leading term in the action  for ``slow'' variables
having non-relativistic Landau-Lifshitz form.
The   solutions
of this reduced sigma model are expected to be  dual to operators
$\Tr (\Phi^{J_1}_1 \Phi^{J_2}_2 + ...)$  {}from the
holomorphic 2-scalar field sector of  the gauge theory.
Indeed, in the next section we shall show
that exactly as in  the undeformed
case \ci{KRUC,KRTS},
 this string action is  the same as the ``Landau-Lifshitz''
action describing macroscopic spin waves  of the
spin chain Hamiltonian  representing the gauge theory
dilatation operator in the 2-scalar sector.

\subsection{A  circular 2-spin  solution \label{class_sol}}


Let us first discuss the case
of  $\sigma_d=0$ where the deformed background
 can be obtained
{}from \adss by means of a T-duality on one circle of
$S^5$, a shift of another angle variable, followed by another
T-duality.
Let us  use the same circular  string
ansatz  as the one  in \cite{ART}
\bea
\la{ans_cir}
\phi_1=\w_1\tau+m_1\sigma\, ,
~~~~~~~~
\phi_2=\w_2\tau+m_2\sigma\, ,
~~~~~~~
\rho_1=\cos\th_0\, ,
~~~~~
\rho_2=\sin\th_0\, ,
\eea
with constant $\th_0$, integer $m_1,m_2$ and
 and $t=\kappa \tau$.
 Then, it is not difficult to check that
if
\begin{eqnarray}
\la{rel_cir}
m_1^2-\w_1^2=m_2^2-\w_2^2\, ,~~~~~~~~\th_0=\frac{\pi}{4}
\end{eqnarray}
we get a solution to the string equations of motion following {}from \rf{A}.
Furthermore, assuming
that both  angular momenta ${\cal J}_1$ and ${\cal J}_2$
corresponding to the
angles $\phi_{1,2}$
are positive, the Virasoro
constraints are satisfied if
\begin{eqnarray}
\vphantom{\bigg|} {\cal J}_1&=& {\cal J}_2\equiv \ha \J \, ,  ~~~~~~~~ m_1=-
m_2\equiv m\, ,
\cr
\w_1 &=& \w_2 = \J + { 1 \ov 2} \tg ( m + \ha \tg\J )
\ .\la{jj} 
\end{eqnarray}
 This  solution  has a smooth limit as
$\tg= \g R^2  \rightarrow 0$.
The corresponding energy  is:
\begin{eqnarray}
\la{en_circ}
E = R^2 \sqrt{{\cal J}^2+(m+ \ha {\tilde\gamma} {\cal
J})^2 }  = \sqrt{ J^2+\l (m+\ha \gamma J)^2 }
\, .
\end{eqnarray}
The undeformed  case  corresponds to $\tg=0$,
i.e.  we conclude that the deformation amounts simply to
 a shift of the string winding number $n$.
Since $m_1=-m_2$,
 then $\phi_1(2\pi) - \phi_2(2\pi) -(\phi_1(0) - \phi_2(0)) =  \g J$
which implies that the role of the
 phase ($\g$) part of the deformation  is to  twist the
boundary conditions of the angle $\phi_1 - \phi_2$ of $S^3$
in the $\sigma$ direction
by $\gamma J={\tilde\gamma}{\cal J}$.

Let us  note also that the expansion of the energy (\ref{en_circ})
\be
E= J[ 1 + \ha  \tl (m+\ha \gamma J)^2  + ... ]
\ee
provides an example of the expected regular expansion
 (\ref{en});
in particular, at the ``1-loop'' (order $\l$)  level we get only
the  terms of order $\g J$ and $(\g J)^2$.

\bigskip


It is easy to generalize the above solution to
the  case of the background  with  $\sigma_d\neq 0$.
We choose the  same circular string
ansatz (\ref{ans_cir})
with constant $\th_0$, and get the same relations (\ref{rel_cir}).
The
Virasoro constraints are satisfied
if again ${\cal J}_1= {\cal J}_2\, , \ m_1=-
m_2$, and
\begin{eqnarray}
\w_1 =\w_2=
 \frac{ {\cal J} +  \tilde{\gamma} ( m + \ha \tg \J)
+ \ha \tilde{\sigma}^2 {\cal J}
}{\sqrt{1+{ 1 \ov 4} \tilde{\sigma}^2}}
\, . \la{soli}
\end{eqnarray}
This solution also has a smooth limit
as $\tb= \tg - i \ts \rightarrow 0$.
The string energy now is:
\begin{eqnarray}
E= R^2\sqrt{   {\cal J}^2+(m+\ha {\tilde\gamma} {\cal
J})^2+\four  \tilde{\sigma}^2 {\cal J}^2  } \ . \la{enn}
\end{eqnarray}
We see again  that the phase of the deformation $\bq$  twists
the boundary condition  of $\phi_1 - \phi_2$ in the $\sigma$
direction by $\tilde\gamma {\cal J}$;
 and we will see below that the $\ts^2$ term
 comes {}from a
 potential term in the reduced (``Landau-Lifshitz'')
sigma model action.
This expression is also  consistent with  (\ref{en}).
 The energy can be also rewritten
as
\begin{eqnarray}
E
=
R^2\sqrt{  {\cal J}^2+|m+\ha  {\tilde\b}{\cal J}|^2}  =
J\sqrt{ 1
 +\tl |m+\ha \b J|^2} \, . \la{eee}
\end{eqnarray}
This expression  for the string energy suggests connection
 with the $SL(2)$  duality transformation
used to obtain the deformed background.



\subsection{``Fast string'' expansion  of the sigma model
  action}

Let us now study   the structure of the
string sigma model action in the
``2-spin'' sector  in the limit of large total spin.
The discussion follows closely the one in the undeformed case
\ci{KRUC,KRTS,KT}.
 Setting
\be\la{var}
\phi_1=\zeta +\eta\ ,
~~~~~~~~
\phi_2=\zeta-\eta\ ,
~~~~~~~
\rho_1=\cos\th\ ,
~~~~~
\rho_2=\sin\th\ , \ \ \ \ \ \rho_3=0 \ ,
\ee
we may
 treat $\z$  (which is the  analog of the angle
$\psi$ in the 2-spin case
and whose conjugate momentum is the total spin
$J=J_1+J_2$) as a ``fast'' angular variable
while $\eta$ and $\th$ will be ``slow'' variables
whose time evolution  will be suppressed.
Then the relevant part of the bosonic  sigma model Lagrangian
\rf{A} in the background \rf{metrgen},\rf{bfield}
can be written as (see \rf{a1},\rf{gh})
\begin{eqnarray}
{\cal L}=&-&\ha R^2
\sqrt{-h}h^{\alpha\beta} H^{1/2} \Big[ -
\partial_\alpha t\partial_\beta t
+
\partial_\alpha \th\partial_\beta \th
+G\left( \partial_\alpha \zeta \partial_\beta \zeta+
 \partial_\alpha \eta\partial_\beta \eta   +
2\cos 2\th\   \partial_\alpha \z\partial_\beta \eta     \right)
\Big]\cr
&- & \ha  R^2 {\tilde\gamma}\, G\sin^22\th\
\epsilon^{\alpha\beta}\partial_\alpha\z\partial_\beta\eta
\ .
\end{eqnarray}
To implement the uniform gauge fixing one may either
consider the  phase-space action and fix $p_\zeta=$const
or, equivalently,  first do T-duality (i.e. 2-d duality)
 in $\z$ direction
in the above Lagrangian and gauge-fix $\td \zeta = J  \sigma$.
The Lagrangian  after T-duality takes the form
\begin{eqnarray}
{\cal L}=&-&\ha \sqrt{-h}h^{\alpha\beta} {R^2}{H^{-1/2}} \Bigg[~~
\frac{\partial_\alpha {\tilde \zeta}\partial_\beta {\tilde \zeta}}
{R^4G}
+
 \frac{{\tilde\gamma}}{R^2}\,\sin^22\th\  {\partial_\alpha
{\tilde \z}\partial_\beta \eta}
\cr &&~~~~~~~
 +H \left( - \partial_\alpha t\partial_\beta t +
\partial_\alpha \th\partial_\beta \th\right)
+\sin^22\th\ \partial_\alpha \eta\partial_\beta \eta \Bigg]\cr
&& ~~~~~~~~ -\cos 2\th\, \epsilon^{\alpha\beta}\,\partial_\alpha{\tilde
\z}\partial_\beta\eta \label{Tdual_action}
\end{eqnarray}
Note that the resulting
 ``Wess-Zumino'' term is identical to the one in the absence of the
deformation.

Imposing now the gauge
(here $\tau$
and $\s$ are world-sheet coordinates)
\be t=\tau \ , \ \ \ \ \ \ \tilde{\z} = J\s \ ,\ \ \ \
\ \ \ \ \ \  J = R^2 \J = \sql \J \ ,  \ee
 and solving for  the
world sheet metric,
we find:
\begin{eqnarray} \la{laa}
{\cal L}&=&   J \cos 2\th\, \,\dot \eta
             - \sqrt{-\det h}\ ,
\end{eqnarray}
where
\begin{eqnarray}
-\det h&=&{R^4}\bigg[~\,
\left(H^{-1} {\sin^22\th} \, {\dot
\eta}\left(\eta'+  \textstyle{\frac{1}{2}}{\tilde\gamma}
{\cal J}\right)  + {\dot\th}\th'\right)^2 \nonumber
\\
&+&
\left(1-{\dot\th}\,{}^2-  H^{-1} {\sin^22\th}\,{\dot\eta}{}^2\right)
\left({\cal J}^2
+H\, \th'{}^2+
\sin^22\th\  | \eta'+ \textstyle{\frac{1}{2}}{\tilde\b}{\cal J} |^2  \right)
\bigg] \ , \la{dett}
\end{eqnarray}
where
\be \la{ji} | \eta'+ \textstyle{\frac{1}{2}}{\tilde\b}{\cal J} |^2
=  ( \eta'+ \textstyle{\frac{1}{2}}{\tilde\g}{\cal J} )^2
+ \four ({\tilde\s}{\cal J} )^2\ . \ee
The  circular  solution we discussed in the previous subsection
\rf{ans_cir},\rf{soli} corresponds 
to the static solution of this
action:
\be\la{new}
\th={\pi\ov 4}\ , \ \ \ \ \ \ \ \ \eta= m \sigma\ ,  \ \ \ \
\sqrt{-\det h}=
J \sqrt{ 1 + \tl  |m +\textstyle{\frac{1}{2}}{\tilde\b}{\cal J} |^2 }
\ , \ee
and so the Hamiltonian part of \rf{laa}
reproduces the energy  of this solution given
 in  \rf{eee}.

To isolate the sector of ``fast strings''
with regular expansion of the energy in $\tl$
 we should take $J$ to be large, i.e. $\tl= {\l \ov J^2} \to 0$,
and expand in time derivatives
(which,  when  eliminated using
leading-order equations of motion \ci{KRTS},
 will contribute to higher orders in $\tl$).
Then
\be
\sqrt{-\det h}
= J \sqrt{ 1 +  \tl \left[
(1 + \four \ts^2  \sin^22\th )\  \th'{}^2
+ \sin^22\th\  | \eta'+ \textstyle{\frac{1}{2}}{\tilde\b}{\cal J} |^2
\right] +  O(\dot \eta, \dot \th)  }
 \ . \ee
The $\J \to \infty$ expansion or
expansion  in powers of $\tl$  makes sense
assuming the parameter $\tb \J $ is  fixed in this limit,
i.e. $\ts\sim \J^{-1} ,\  \tb \sim \J^{-1} $.
Then to the leading order in $\tl$  the prefactor of the
$ \th'{}^2$ term  contributes simply 1   and we get  for the
resulting Lagrangian
\be
\la{LL}
{\cal L}=   J \left[   \cos 2\th\, \,\dot \eta
      -\ha  \tl
  \left(  \th'{}^2
+ \sin^22\th\  | \eta'+  \textstyle{\frac{1}{2}}{\tilde\b}{\cal J} |^2
\right)  + O(\tl^2 )     \right]  \ . \ee
Thus the real part of the deformation parameter $\beta$ twists
the boundary conditions of the world-sheet field $\eta$ while the
imaginary part adds a nontrivial potential term (cf. \rf{ji}).\foot{
Let us note that even though the Lagrangian  (\ref{LL}) formally
depends  on the real and
imaginary parts of the deformation parameter $\beta$
through $\tg - i \ts$, it is not clear
that the same is true  for  energies of all semiclassical solutions (as
we have seen in the previous subsection,
 for some of them this is still the case, cf. \rf{eee}). Indeed,
the equations of motion following {}from (\ref{LL}) depend separately on
$\tg$ and $\ts$ and this may  continue once their solutions
are plugged into  the Hamiltonian.}
Note that the $\s$-momentum  quantization  condition \ci{KRUC,hl,KT}
is  unchanged from the
$\beta=0$ case: the variations of $\theta$ and $\eta$  under
$\s$-translations remain the same, as are  the momenta conjugate to them,
so that we get $ \int {d\s\ov 2 \pi} \cos 2 \theta\ \del_\s
\eta=$integer.

This action leads to the anisotropic Landau-Lifshitz equations with
twisted boundary conditions for the $\eta$ field
(the anisotropy potential term is present only if  $\ts\not=0$).
 These equations are
known to be integrable \cite{FADT},  and this  suggests that the  string
theory in this  2-spin sector is integrable
(at least  to the leading order in the
$\tl $ expansion).



The action (\ref{LL})  admits a simple classical solution 
  with two
unequal  angular momenta  $J_1,J_2$  that generalizes the 
solution in \rf{new} and should be a  large $\J$ limit 
of a $J_1\not=J_2$  generalization of the solution 
\rf{jj} (i.e. the counterpart of the circular solution of \ci{ART}).
Translating the string ansatz (\ref{ans_cir})
into the variables \rf{var}
 appearing in (\ref{LL}) we find
\begin{equation}
\theta=\theta_0\ , ~~~~~~~~~~~~\eta\equiv\ha (\phi_1-\phi_2)=
w\tau+\ha {\mm}\sigma\ , 
\label{ans_gen}
\end{equation}
where $\mm=m_1-m_2$ is an integer  and $w$
is a constant  which is to be determined by solving 
equations following from \rf{LL}.\foot{It will be the leading-order term 
in the large $\J$ expansion of $\frac{1}{2}(w_1-w_2)$. 
Note that the condition 
$m_1J_1+m_2J_2=0$  following from the Virasoro 
 constraint in the string theory setting,  
 reappears  in the Landau-Lifshitz  context  from 
 the momentum quantization condition 
 $ \ha ( m_1+m_2) = \int {d\s\ov 2 \pi}\del_\s \zeta 
 = -\int {d\s\ov 2 \pi} \cos 2 \theta\ \del_\s \eta $.}
It is  easy to see that the constant $\theta$ 
assumption 
implies that $\eta$
equation of motion is trivially satisfied while the $\theta$ equation
fixes $w$ in terms of $m$ and $\theta_0$. The latter 
is determined in terms of the angular momenta. Indeed, the relation 
between $\eta$ and $\phi_1$ and $\phi_2$ (\ref{ans_gen}) 
implies that the momentum conjugate to $\eta$ is given by the
difference between the conjugate momenta of $\phi_1$ and $\phi_2$,
i.e. $J_1-J_2$. Thus, 
\begin{eqnarray}
J_1-J_2=J\int_0^{2\pi}\frac{d\sigma}{2\pi}\,\cos2\theta_0=J\cos2\theta_0
~~\longrightarrow~~
\sin^2 2\theta_0=4\alpha(1-\alpha)~~{\rm with}~~\alpha\equiv \frac{J_2}{J}~~.
\label{theta0}
\end{eqnarray}
Then  we find that
\begin{equation} 
w=-\frac{1}{4}{\tilde\lambda}(1-2\alpha)\Big|\mm+
{\tilde\beta}{\cal J}\Big|^2~~.
\end{equation}
The equation (\ref{theta0}) also implies that, to leading order in the
${\tilde\lambda}$ expansion, the energy of the
solution (\ref{ans_gen}) as determined from \rf{LL}  is given by 
\begin{equation}
E_0=\frac{1}{8} J {\tilde\lambda}\sin^2 2 \theta_0\ \Big|\mm+
{\tilde\beta} {\cal
J}\Big|^2 =
\frac{{\lambda}}{2J}\alpha(1-\alpha)\Big|\mm+
{\tilde\beta} {\cal J}\Big|^2\ . 
\label{en_gen}
\end{equation}
Note that in the case of $J_1=J_2$ ($\a= \ha$)
 in \rf{new}
we have $\mm= 2m$ and \rf{en_gen} agrees with the leading term 
in the expansion of the square root in \rf{new}.

We conclude that, as in  the undeformed case, the
deformed (``twisted'' and ``anisotropic'') Landau-Lifshitz 
action \rf{LL} has a simple rational solution for generic 
value of $J_1-J_2$.  
We will reproduce  (\ref{en_gen}) in
\S 8.2 from the Bethe ansatz for the corresponding  spin chain.

\bigskip

 One can also compute the first subleading correction to the 2d Hamiltonian
 corresponding to  \rf{laa}
\bea \la{ha}
\HH = J \int^{2\pi}_0 {d \s \ov 2 \pi} \ \H \ , \ \ \ \ \ \ \ \ \ \ \
\H = \H_0 + \H_1 + \H_2 + ... \ , \\
\H_0 = 1 \ , \ \ \ \ \ \ \ \ \ \ \ \ \ \
\H_1=  \ha  \tl
  \left(  \th'{}^2
+ \sin^22\th\  | \eta'+  \textstyle{\frac{1}{2}}{\tilde\b}{\cal J} |^2
\right) \ ,  \la{haha}  \eea
using the same procedure of eliminating time derivatives
with the help of the leading-order equations of motion as in \ci{KRTS}.
One finds then for the $\l^2$ term:
\begin{eqnarray}
 \H_2=- {{\textstyle {1 \ov 8}}}  \tl^2 &\bigg[&
     \left(
 \th'\,{}^2
  +\ \sin^22\th \ \big| \eta'+\textstyle{\frac{1}{2}}
 {\tilde\beta}{\cal J} \big|^2
                           \right)^2
  -
( {\tilde\sigma}{\cal J})^2 \sin^22\th\,\th'\,{}^2\nonumber
\\
& + &\left[
4\cos 2\th\ \th' ( \eta'+ \textstyle{\frac{1}{2}} {\tilde\beta}{\cal J} )
 +\sin 2\th\ \eta''
\right]^2
+
 \left(
\th''
 - \sin 4\th \
\big| \eta'+\textstyle{\frac{1}{2}}{\tilde\beta}{\cal J} \big|^2
                        \right)^2
\Bigg]
\label{2loopLL}
\end{eqnarray}
Note  that this correction to the Landau-Lifshitz action
depends separately on the real and imaginary parts of the deformation
parameter $\b$. This can be  traced to the fact that while
in the leading order action the deformation parameter enters only as a
(complex) twist in the boundary conditions of the $\eta$ coordinate,
this is not true at the level of the equations of motion.

\renewcommand{\theequation}{5.\arabic{equation}}
 \setcounter{equation}{0}

\section{Coherent state action for $\beta$-deformed ${\cal N}=4$ SYM
\label{coh_state_action}}

To try to establish the AdS/CFT correspondence in the present deformed case
let us now turn to gauge theory and  address the question of
which gauge-theory operators correspond to the fast-spinning strings
discussed in the previous section and which are their 
anomalous dimensions.

The first   important issue   is whether in the
deformed theory there exist subsets  of relevant  single-trace
scalar operators which are
closed under the renormalization group flow. A brief look at the
Feynman rules in the $\b$-deformed theory combined with the fact that
the deformation preserves the three Cartan generators of the $SO(6)$
$R$-symmetry group of ${\cal N}=4$ SYM leads one  to the conclusion
that at the 1-loop level all sectors closed in the original undeformed theory
remain closed in the presence of the deformation. Given
that the deformation does not introduce new types of  interactions,
this observation may be extended to higher loops as well. In the
following we will be concerned in detail only with the operators
built out of the  two holomorphic scalar fields, i.e. $\Tr (\Phi_1^{J_1} \Phi_2^{J_2})+ ...$.
We will call this
{\it  $su(2)_\b$ or ``2-spin''} sector,  by analogy with the
$su(2)$ sector of the  ${\cal N}=4$ SYM to which it reduces
when the $\b$-deformation is switched off.
The sector of operators  built out of the  three holomorphic scalar
fields will be called the $su(3)_\b$ sector.

The dilatation operator of any field theory with matrix degrees of
freedom can be represented as a Hamiltonian of a spin chain
acting on states in the spin chain Hilbert space. Indeed,
for  single-trace composite operators
the matrix structure of the field theory defines a 1-dimensional
lattice structure
can be put into one-to-one correspondence with the lattice structure
of the spin chain states. Then any operator acting on
field theory composite operators has a (unique, up to
symmetries)
 representation  as an operator acting on the spin chain states.
In the planar limit
all such spin-chain operators are local, i.e.  act on
adjacent spin chain sites.
An example is provided by  the dilatation operator (see, e.g., \ci{B,D}).
Constructed in perturbation theory, the $L$-loop (order $\l^L$) term in this  operator
acts on at most  $L+1$ spin chain sites.
 At 1-loop this
becomes a nearest-neighbour interaction,  and given such
interaction term  it turns out to be very easy to engineer a field
theory which has it as its dilatation operator in some sector.

An  interesting property of the spin chain Hamiltonians
in  various sectors of  ${\cal
N}=4$ SYM theory is that they are integrable.
Remarkably, it  turns out \cite{RR,BECH}
 that the spin chain appearing in the $su(2)_\beta$ and
$su(3)_\beta$ sectors of the $\beta$-deformed ${\cal N}=4$ SYM theory are
also integrable for an  arbitrary complex  (in $su(2)_\beta$   case)
and real  (in $su(3)_\beta$   case)  deformation parameter,
respectively.

\bigskip

Before embarking on a detailed  analysis of  the properties of the
spin chain for the $su(2)_\beta$ sector  in this section we will
first convince ourselves,  following \cite{KRUC,KRTS},  that its
Hamiltonian \cite{RR} can  indeed  be related to the world-sheet
string theory in the $\beta$-deformed background. Explicitly,
 we shall construct the effective 2d action for  the low energy
``macroscopic'' excitations of the spin chain in the special scaling
limit when the length of the spin chain is taken to be large
with $\tl={\lambda\ov J^2}$,  $J_1/J_2$ {\it and $\b J$} (cf. \rf{tb})
 kept fixed,
\be \la{sca}
L\equiv J\rightarrow\infty\ , \ \ \ \ \ \ \
{\tilde\lambda}={\lambda\ov J^2}={\rm fixed} \ ,\ \ \ \ \ \ \
\b J = {\tilde  \b \ov  \sqrt{\tl}} = {\rm fixed}  \ ,  \ee
and  show that it matches   the string
sigma model action \rf{LL} expanded in the same limit.
As in the undeformed case,
this  implies the remarkable matching between the leading terms
in the corresponding string energies
and the 1-loop gauge-theory anomalous dimensions.

\bigskip

In  Appendix A we shall review the structure
of the  spin chain \cite{RR} for a
3-parameter family of deformations of ${\cal N}=4$ SYM which
contains as a special case the present 
 $\beta$-deformed superconformal theory. There are
several ways of writing the corresponding
spin chain Hamiltonian which are
unitary equivalent. Here we find it convenient to present  the Hamiltonian
(\ref{hamiltonian}) in the following form
\begin{eqnarray}
&&{\HH}= \frac{|h|^2\lambda}{(4\pi)^2} \sum_{l=1}^J \Bigg[ -
\left(\bfsigma^x_l\otimes\bfsigma^x_{l+1} +
\bfsigma^y_l\otimes\bfsigma^y_{l+1}+\bfsigma^z_l\otimes\bfsigma^z_{l+1}
- \id_l\otimes\id_{l+1}\right)  \cr &&~~~~~~~~~~~~~~~~~~~~~~~~
+(1-\cosh 2\pi\k_d)
\left(\bfsigma^z_l\otimes\bfsigma^z_{l+1} -
\id_l\otimes\id_{l+1}\right)\la{H}\\
&&\!\!\!\!\!\!\!\!\!\!\! {
+\ (1-\cos2\pi\beta_d)\ (\bfsigma^x_l\otimes\bfsigma^x_{l+1} +
\bfsigma^y_l\otimes\bfsigma^y_{l+1} )  +  \sin2\pi\beta_d\
(\bfsigma_l^x\otimes\bfsigma_{l+1}^y-\bfsigma_l^y\otimes
\bfsigma_{l+1}^x)} \Bigg]~~.\nonumber
\end{eqnarray}
Here $\bfsigma^i$ are Pauli matrices,
and    $\b_d$ and $\k_d$ are the deformation parameters {}from \rf{kkk}.
This is a  Hamiltonian of 
a ferromagnetic  XXZ spin chain \ci{Fad} with broken
parity invariance.\footnote{The parity-invariant XXZ spin chain with
 $\cosh
2\pi\kappa_d=3$ appears as a 1-loop
  dilatation operator of  ${\cal N}=2$ SYM theory
\cite{DVTA}.} 
In the conformal limit the coefficient $|h|^2$ is related to the
deformation parameter by the condition (\ref{confcond}), 
i.e. $h$ in general is not equal to 1 if $\beta$ is complex.
 However, for the
purpose of comparison with string theory we may  set 
 here  $|h|^2=1$.
 Indeed, 
as discussed in section 3, we are interested in the  limit 
when $J$ (angular momentum or  the spin chain length)
is large while $\beta J$ (and $\l/J^2$)  stays finite, i.e. 
the deformation parameter  $\beta $ is scaled to zero or $|e^{i\pi \beta}|\to 1$.
We will thus ignore the factor $|h|^2$ in $H$  in the 
following.\footnote{Ignoring the factor of $|h|^2$ in $H$ will also be
consistent in section \ref{subleading} where we will study $1/J$
corrections. Indeed, the equation (\ref{confcond}) implies that
$|h|^2=1+{\cal O}(1/J^2)$ and thus will be subleading also at that
order.} 

 An alternative spin chain (with
parity-invariant Hamiltonian) was discussed for the same theory  in
\cite{BECH}; it is obtained {}from the one above by setting
$\beta_d=0$ in $\HH$ while at the same time twisting
the boundary conditions by $\beta_d$.
The first line in
(\ref{H}) is the Heisenberg XXX spin chain Hamiltonian
of  the $su(2)$ sector of
the ${\cal N}=4$ SYM \ci{MZ}
while the other terms arise  due to
the $\beta$-deformation. It is possible to remove the last term by
performing a position-dependent unitary transformation
\cite{BECH}.

The construction of the corresponding effective
action in the  continuum limit proceeds in the standard
way (see, e.g., \ci{SA}).
The first step is to rewrite  the spin chain  path integral as
an integral over the overcomplete basis of coherent states
\begin{equation}
Z=\int[dn]\ e^{iS[n]}~\ , ~~~~~~~~~~~~~~~~~~~~ S=\int dt\, \left(L_{WZ} -
 \langle n|\HH |n\rangle \right) \ , \la{coh}
\end{equation}
where the first term $L_{WZ}$ is the analog of the $p{\dot q}$
term in a phase space action coming {}from the Berry phase
$\langle n|i \del_t |n\rangle $.

In general,  one chooses coherent states  based on
symmetry of the discrete Hamiltonian one is interested in.\foot{
The coherent states usually parametrize the coset $G/H$ where
$G$ is a symmetry of the Hamiltonian and  $H$  is a symmetry
of the vacuum.}
 In our present
case the (hidden) symmetry of the  chain \rf{H} is
$\widehat{SU(2)_{q_d}}$ -- the central extension of the quantum
deformation of $su(2)$ (where $q_d$ is defined in \rf{vev}), so
one  might be tempted to use the coherent
state of the fundamental representation of this group.

It turns out to be sufficient for our purposes
to use the same standard $SU(2)$ coherent states  parametrized
by a unit vector $\bn$ at each site $l$ as in the undeformed case,
\begin{eqnarray}
&&\langle n|\bfsigma_i|n\rangle = n_i ~ ,\\
 &&
\sum^3_{i=1} n^2_i =1 ~, ~~~~~~~ {\bf n}=(\sin 2\th\cos 2\eta,
\sin 2\th\sin 2\eta, \cos 2\th)\ .
\label{coh_st}
\end{eqnarray}
Then the Wess-Zumino
term  in the resulting coherent-state action will be the same
as in the undeformed case, i.e.
\begin{eqnarray}
L_{WZ}= \sum_{l=1}^J  {\cal L}_{WZ} (\bn_l, \dot \bn_l)
\ , \ \ \ \ \ \ \ \
{\cal L}_{WZ}=
-\frac{1}{2}\int_0^1ds\,\epsilon_{ijk}n^i\partial_s n^j\partial_t n^k
=     \cos 2 \th\  \dot \eta \ ,\la{wz}
\end{eqnarray}
where, as usual,   $s$ is an auxiliary variable needed to put
$L_{WZ}$ in a manifestly $SU(2)$-invariant form.



Computing the expectation value of the Hamiltonian \rf{H}
we would like to be able to take a continuum limit
and define a semiclassical approximation for the resulting path integral,
dominated by the classical action.
For this to be possible, we need, as in the undeformed case \ci{KRUC,KRTS},
 to
be able to  put the action in \rf{coh} in the form with  the
factor $J$ appearing in front of it, so that $1/J$  plays the role of an
effective Planck's constant.
Introducing the  1-d  lattice spacing
$a= {2\pi\ov J}$ and defining $\bn(\s)$ with $\bn_l = \bn(a l)$
($l=1,2...,J$)  we conclude that  we indeed need to fix $\tl $ and $\b J$
or $\tilde \b$  in \rf{sca} (i.e. to assume $\b \sim a$)
 to be able to define this scaling limit.

We then get for the continuous limit of the  expectation
values of  different terms in \rf{H}:

\noindent -- the first line in \rf{H} gives the same as
 the  continuum limit of the
Heisenberg XXX Hamiltonian:
\begin{equation}
\frac{1}{2}a^2\, ({\bf n}')^2+\frac{1}{24}a^4 \, ({\bf n}'')^2 + ...=
2a^2\left(\th'{}^2+\sin^22\th\,\eta'\,{}^2\right)
+ O(a^4) \ ,  \label{vev0} \label{l1}
\end{equation}

\noindent -- the second line  in \rf{H}:
\begin{eqnarray}
(n_z^2 -1)-\frac{1}{2}a^2 (n_z')^2 + ... =-\sin^22\th
+ O(a^2)  \label{l2}
\end{eqnarray}

\noindent -- the first term on the third line in \rf{H}:
\begin{eqnarray}
(n_x)^2+(n_y)^2-\frac{1}{2}a^2 \left[(n_x')^2+(n_y')^2\right] + ...=
\sin^22\th  + O(a^2)  \ ,
\label{l3_1}
\end{eqnarray}

\noindent -- the second term on the third line in \rf{H}:
\begin{eqnarray}
a\left(n_xn_y'-n_yn_x'\right)-\frac{1}{2}a^3 n_x'n_y''+ ...= 2a \sin^2
2\th\,\ \eta'  + O(a^3)    \ . \label{l3_2}
\end{eqnarray}

Combining the leading order terms in (\ref{l1}), (\ref{l2}),
(\ref{l3_1}) and (\ref{l3_2}) in the continuum limit  $a= {2 \pi \ov J} \to 0$ we get
\be \la{gge}
 \langle n|\HH |n\rangle \ = J \int^{2\pi}_0 {d \s \ov 2\pi} \ \H_1+... \ ,
\ee
\be\la{kou}
\H_1 =   {\l \ov 2J^2}  \left(\th'{}^2+\sin^22\th\,\eta'\,{}^2\right)
+   {\l\ov 16 \pi^2}  ( \cosh 2 \pi \k_d - \cos 2 \pi \b_d) \sin^2 2\th
+ {\l\ov 4 \pi J}  \sin 2 \pi \b_d \ \sin^2 2\th\,\eta'  \ .
\ee
It is easy to see that, as in the undeformed case,
${\bf n}_0 =(0,0,1)$ corresponds  again  to  the vacuum
of the Hamiltonian \rf{H} represented by the BPS operator $\Tr
\Phi^J$. Indeed, $\HH$ vanishes  for $\th=0, \eta=0$.

As follows {}from \rf{kou},
to  be able  to define an effective action  for  which
 $\langle n|\HH |n\rangle $ will  be small at large $J$
we  thus need  to assume  that $\k_d$ and $\b_d$ should  be taken to zero
with $J \k_d$ and $J \b_d$  kept fixed.\foot{If $\b_d$ and $\k_d$ are kept finite
at large $J$ as would be  for {\it discrete} values of the deformation parameters,
(as, e.g., in the case of $\N=2$ SYM discussed in \ci{DVTA})
the energies of spin-chain states   will get large
shifts proportional to $J$  coming {}from $\sin^2\th$ potential
term  and matching onto string states
will not be possible.}
Then,
\bea
\H_1&=&   {\l \ov 2J^2}  \left[\th'{}^2+\sin^22\th\,
\left(\eta' + \ha \b_d J \right)^2 
+   \four (\k_d J)^2 \sin^22\th\ \right]  \nonumber \\
 &=& \ha{{\tilde \lambda}} \left(
\th'\,{}^2+\sin^22\th\ \Big|\eta' + \textstyle{\frac{1}{2}}
\beta J \Big|^2\right) \ .\la{kop}
\eea
This is exactly the same as the anisotropic Landau-Lifshitz
 Hamiltonian \rf{haha}
we found  above on the string theory
side.\foot{Note that,  as was   expected {}from the string-side discussion, we
did not obtain any correction to the coefficient of $\th'^2$.}
Also,  the time-derivative dependent WZ term in the resulting
action  is the same as in the undeformed case, i.e.
the same  as in \rf{LL}, so the two
effective actions are exactly the same.

The potential $\sin^2 2\th$ term has its absolute minimum at  BPS state
${\bf n}_0 =(0,0,1)$ or $\th=0, \eta=0$.
The spectrum of small fluctuations
 near  this  BPS vacuum  can be  found {}from the 
 Landau-Lifshitz action
\rf{LL}  by expanding to quadratic order in the  $n_1$ and $n_2$
components of $n_i$ (the expansion in the fluctuations of the 
angles 
$\th, \eta$ is singular). 
One  finds  the analog of the  BMN  spectrum
with the energy scaling as
\be\la{bmn}
E_n =   {\l \ov 2 J^2}  \Big|n +  \beta J \Big|^2
\ . \ee
Note that there is no 1/2 factor in front of $\beta J$
in contrast to what one could naively expect from 
\rf{kop}.
We shall obtain the same spectrum
directly {}from the Bethe ansatz for the spin chain
in the next section.


\renewcommand{\theequation}{6.\arabic{equation}}
 \setcounter{equation}{0}

\section{Bethe Ansatz for the spectrum of anomalous dimensions
of  $\beta$-deformed $\N=4$ SYM in the
2-spin sector \label{spin_chain_stuff}}


The Hamiltonian (\ref{H}) representing  the one-loop dilatation
operator in the $su(2)_\b$ sector is integrable via the Bethe
ansatz for all values of its parameters;
the Bethe equations and the cyclicity
condition\footnote{In the expression below we correct 
 a misprint in \cite{RR}.} are  (in this section we use the notation
$L\equiv J, \ M\equiv J-J_1= J_2$)
\bea
\nonumber e^{-2\pi i \b_d L} \left[ {\sin(2\pi\k_d (\nu_k +
i/2))\ov \sin(2\pi\k_d (\nu_k - i/2))}\right]^L &\=&
\prod_{\stackrel{j=1}{j\ne k}}^M {\sin(2\pi\k_d (\nu_k-\nu_j +
i))\ov \sin(2\pi\k_d (\nu_k-\nu_j - i))}
\, ;\\
e^{-2\pi i \b_d M}\prod_{k=1}^M {\sin(2\pi\k_d (\nu_k + i/2))\ov
\sin(2\pi\k_d (\nu_k - i/2))} &\=&1\, , \la{BE} \eea
where $\nu_k$ are rapidities of elementary excitations.
 The energy of
a general state and the anomalous conformal dimension of the
corresponding field theory operator
$\Tr (\Phi_1^{L-M} \Phi_2^M ) + ...$
is given by
\begin{equation}
E(\{\nu_k\})\=\sum_{k=1}^M\,\epsilon_k~~~~{\rm with}~~~~
\epsilon_k\=\frac{\lambda}{8\pi^2}\,{\sinh^2(2\pi \k_d)\ov
\sin(2\pi\k_d (\nu_k + i/2))\sin(2\pi\k_d (\nu_k - i/2))}\,.
\label{energ_parts}
\end{equation}
The only difference of this  Bethe ansatz {}from the usual one for the
XXZ spin chain is in the presence of the additional
$\b_d$-dependent phase factors in the l.h.s. of the equations. The
$\beta_d$ dependence  factors out of the Bethe equations and the
cyclicity condition  for generic $\k_d$. This is consistent
with the treatment  of the phase ($\b_d$) of the deformation
\rf{vev} in \cite{BECH}.

\subsection{One-loop Long Chain}

The Bethe equations can be used to compute anomalous dimensions of
primary operators in the $su(2)_\b$ sector  for finite $L$
just as it was done for the $su(2)$ sector of ${\cal N}=4$ SYM
\cite{MZ}. For comparison with semiclassical
string theory we are
interested in studying long spin chains, $L\gg1$. There are two
different large $L$ limits: $(i)$ the ``BMN limit'':\ $L\to\infty$ with
the number of excitations $M$ kept fixed,\
 and $(ii)$ the
``thermodynamic limit'':\  $L\to\infty$  with the ratio $M/L$ kept
fixed. The BMN limit is used  to compute conformal
dimensions of near-BPS  (BMN) operators \cite{BMN}, and the thermodynamic
limit is used to find dimensions of operators dual to multi-spin
string states \cite{FT}.

Below we will use the above Bethe equations to derive the anomalous dimensions
of BMN-type operators and the integral Bethe equations arising in
the thermodynamic limit in the $\beta$-deformed theory. This  will
give us some understanding of the relation between (\ref{BE})
and the Bethe equations in the $su(2)$ sector of
undeformed ${\cal N}=4$ SYM.
{}From the  gauge  theory standpoint,  the anomalous dimensions of the
BMN-type operators  were  computed in \cite{NIPR} using the
technique of \cite{SAZA}.
We will also derive  integral Bethe equations  appearing in the
thermodynamic limit  which later will be
compared with the ``string Bethe equations''  which we will
find {}from the string sigma model  in the next section.

\

Since the details of the complex deformation parameter  are
somewhat more involved, we will begin by setting its imaginary
part to zero and restore it later on.

\

\noindent $\bullet$ {\bf Case of real $\b$}:
According to \rf{kkk}
(see also Appendix A)
this deformation corresponds to
\begin{equation}
\k_d = 0~~.
\end{equation}
In this limit the Bethe equations, the cyclicity condition and the
energy of an individual excitation simplify considerably: \bea
e^{-2\pi i\beta_d L}\left(\frac{\nu_k+i/2}{\nu_k-i/2}\right)^L&\=&
\prod_{j\ne k=1}^M \frac{\nu_k-\nu_j+i}{\nu_k-\nu_j-i}
~~~~;~~~~~~~~
e^{-2\pi i\beta_d M}\prod_{k=1}^M\frac{\nu_k+i/2}{\nu_k-i/2}\=1\,,
\label{tw_Bethe_cyclic}
\\ \la{eenn}
\epsilon_k&\=&\frac{\lambda}{8\pi^2}\,{1\ov \nu_k^2 + 1/4}\,. \eea
We see that, up to the phase factors, these Bethe ansatz equations coincide
with the usual ones for the XXX spin chain that  describes
the $su(2)$ subsector of the $\N=4$ SYM. Moreover, the expression
of the energy in terms of the rapidities $\nu_k$ remains the same.
It is clear, therefore, {}from (\ref{tw_Bethe_cyclic}) that to
describe long spin chains we should rescale the rapidities
\be \la{resc}
\nu_k = L x_k \ , \ee
 and
expand the equations in powers of $1/L$. Then these Bethe equations
and the cyclicity condition reduce to the equations
associated to the XXX
 Heisenberg chain  \ci{BMSZ}
with the mode and momentum number shifted in a
$\beta_d$ -dependent way:
\begin{eqnarray}
\frac{1}{x_k} &\=& \frac{2}{L}\sum_{\stackrel{j=1}{j\ne k}}^M
\frac{1}{x_k-x_j}+2\pi(n_k+\beta_d L)
~;~~~~~~~~
\sum_{k=1}^M  \frac{1}{x_k}=2\pi (m+\beta_dM)L~~.
\label{sh_T_lim}\\
\epsilon_k&\=&\frac{\lambda}{8\pi^2L^2}\,{1\ov x^2_k} \,.
 \label{ener_p}
\end{eqnarray}
It is easy to verify that the existence of  both the BMN and the
thermodynamic limits requires that $\beta_d L$
should remain finite as $L\rightarrow\infty$, in a nice agreement with
the discussions in \S \ref{limpar} and 5.

The equation (\ref{sh_T_lim}) implies that {\it all} the solutions of
the Bethe equations in the large $L$ limits relevant for    the undeformed
 ${\cal
N}=4$ SYM theory  can be easily translated into  the
theory deformed by  a real $\beta$.

\

A simple  illustrative example  is that of the  BMN-type operators --
operators for which the number of excitations is kept finite as
the length of the chain is taken to infinity. In this case the
first term on the right hand side in the first equation
(\ref{sh_T_lim}) becomes subleading and the scaled rapidities
$x_k$ are
\begin{equation}
x_k=\frac{1}{2\pi (n_k +\beta_d L)} \label{rap_BMN}
\end{equation}
while, in spite of the deformation, the cyclicity constraint
becomes the same one as in the $su(2)$ sector of ${\cal N}=4$ SYM
\begin{equation}\la{mem}
\sum_{k=1}^M n_k  = 0~~.
\end{equation}
As in that case, since the mode numbers $n_i$ are fixed and 
$M/L\rightarrow 0$ as $L\rightarrow\infty$,  
the second equation \rf{sh_T_lim} implies that the momentum number
$m$ vanishes.
Combining all the ingredients we find  the 
anomalous dimension of BMN-type operators
\begin{equation}
E(n_1,\dots, n_M)= \frac{\l}{2L^2}  \sum_{k=1}^M \left(
{n_k}+\beta_d L \right)^2 =
\frac{1}{2} \lambda \sum_{k=1}^M \left(
\frac{n_k}{L}+\beta_d\right)^2~~. \label{energ_parts_phase}
\end{equation}
This agrees with the leading order term in the
expansion
of the energies of the string states in the plane wave limit of the
deformed background  found  in \cite{LUMA} (see also \cite{NIPR})
and confirms the
identification of the gauge theory
parameter $\beta_d$ with the string theory quantity
${\beta}\,\big|_{\sigma_d=0}$.

\

To derive the integral Bethe equations arising in the
thermodynamic limit of  the $\beta$-deformed theory we proceed in
the same way as in the $\b=0$ case treated  in  \ci{BMSZ}.
Introducing the distribution density
\bea
\la{distr_dens}
{1\ov L}\sum_{k=1}^M\, f(x_k)\=\int_C {\rm d}\xi\,
\r(\xi)f(\xi)\, ;\qquad
\r(\xi)\= {1\ov L}\sum_{k=1}^M\, \delta(\xi - x_k)\, ,
\eea
with the support on a discrete union of $n$ contours
$C = C_1\cup C_2\dots \cup C_n$, we obtain the integral Bethe
equations
\bea \la{gfinteq}
2\, \pint_C {\rm d}\xi\,
{\r(\xi)\ov x-\xi}&\=& {1\ov x}\- 2\pi\left(n_k+\b_d L\right)\,,\quad
\ \ \ x\in C_k\,,\\
\la{gfdens1}
\int_C {\rm d}\xi\, {\r(\xi)\ov \xi} &\=&
2\pi\left(m+ \b_d M \right)\,;\qquad \int_C {\rm d}\xi\, \r(\xi)\=
{M\ov L} \,.
\eea
The energy of a general state and the anomalous
 dimension of the corresponding field theory operator are
expressed in terms of  the density as follows
\bea \la{gfdens3}
E = \frac{\lambda}{8\pi^2L^2}\,\int_C {\rm d}\xi\, {\r(\xi)\ov \xi^2} \,.
\eea
The equations (\ref{gfinteq})-(\ref{gfdens3}) differ {}from
the integral Bethe equations of the $su(2)$ sector of $\N = 4$ SYM \ci{BMSZ}
{\it only by the shifts of the mode numbers and the
momentum number}.
Therefore, {\it all} the solutions found in the  ${\cal N}=4$ SYM  case
 can be easily translated into  the theory deformed
by  a real $\beta$. In \cite{KMMZ} all  solutions of the above integral
Bethe equation in the $\b=0$ theory
have been constructed in terms of (noncompact) Riemann surfaces
endowed with meromorphic differentials with integer periods. These
periods over compact cycles are given by differences of mode
numbers and as such  are  {\it invariant} under the shift in
(\ref{gfinteq}). The only period sensitive to this shift is the
one over the noncompact cycle of the Riemann surface.

\bigskip

\noindent $\bullet$ {\bf Case of complex  $\b$}:
While qualitatively similar,
this case is a little bit more involved because we now have to
also keep track of $\k_d$. It is not difficult to see {}from
(\ref{BE}) that,  for large $L$, $\k_d$ gives a nontrivial
contribution to the Bethe equations if $\k_d L$ is kept fixed in
the large $L$ limit.
This implies in particular that
 ${\tilde q}_d=q_d^L= \exp(-2\pi{\k_d}{ L})$
and $\td \s$ in (\ref{ident3}),\rf{tb}
is kept finite on the string theory side.

To analyze the Bethe equation in this limit we find it convenient to
make the following change of the rapidities $\nu_k$
\begin{equation}
\la{change_nu}
\tan(2\pi \k_d\,\nu_k)\= 2\tanh(\pi\k_d)\ L\, x_k\,.
\end{equation}
If $\k_d\to 0$ then this redefinition becomes  equivalent to
the rescaling  \rf{resc}  used to analyze the real $\b$ case.

Written in terms of  the new parameters
 $x_k$, the Bethe equations (\ref{BE}) and
(\ref{energ_parts}) take the following
form
\bea \nonumber
e^{-2\pi i \b_d L} \left[ {L x_k+i/2\ov L x_k-i/2}\right]^L
&\=&
\prod_{\stackrel{j=1}{j\ne k}}^M
{x_k-x_j+{i\ov L}{\tanh(2\pi \k_d)\ov 2\tanh(\pi
\k_d)}(1+4\tanh^2(\pi\k_d)L^2x_kx_j)\ov x_k-x_j-{i\ov
L}{\tanh(2\pi \k_d)\ov 2\tanh(\pi
\k_d)}(1+4\tanh^2(\pi\k_d)L^2x_kx_j)} \, ;\\
e^{-2\pi i \b_d M}\prod_{k=1}^M {L x_k+i/2\ov
L x_k-i/2} &\=&1\, ; \la{BEk}\\
E(\{\nu_k\})\=\sum_{k=1}^M\,\epsilon_k&{\rm with}&~~
\epsilon_k\=\frac{\lambda}{8\pi^2}\,{\cosh^2(2\pi \k_d)\left(1+
4\tanh^2(\pi\k_d)L^2x_k^2\right)\ov L^2x_k^2+1/4}\,.
\label{energ_pa}
\eea
Now the expansion of these equations in
powers of $1/L$ is straightforward.
Taking the logarithm,
 we obtain the following
system which also describes subleading $1/L$ corrections to the
Bethe equations\footnote{There exists a potential for anomalies
due to roots with spacing of order $1/L$ \cite{BTZ,HLO}. We will return to
 this
in \S \ref{subleading}.}
\begin{eqnarray}
\frac{1}{x_k} &=& \frac{2}{L}\sum_{\stackrel{j=1}{j\ne k}}^M
\frac{1+(2\pi \k_d L)^2x_kx_j}{x_k-x_j}+2\pi(n_k+\beta_d L)
~;~~~~~~
\sum_{k=1}^M  \frac{1}{x_k}=2\pi (m+\beta_dM)L~;~~
\label{BE_large_L}\\
\epsilon_k&\=&\frac{\lambda}{8\pi^2L^2}\,\left[{1\ov x^2_k}+
(2\pi\k_d L)^2\right] \= \frac{\lambda}{8\pi^2L^2}\,{1\ov x^2_k}\+
{1\ov 2}\l\k_d^2
\,.
 \label{ener_pa}
\end{eqnarray}
Recall that both $\b_d L$ and $\k_dL$ should be  kept fixed in the large
$L$ limit.

We see that the energies of individual excitations differ {}from the
real $\b$ case only by a constant shift $\ha \l\k_d^2$. Moreover, in
the BMN limit ($L\rightarrow\infty$ with $M/L\rightarrow 0$)
the equations (\ref{BE_large_L}) coincide with the equations
(\ref{sh_T_lim}) describing the BMN limit in the real $\b$ case.
Therefore, the anomalous dimensions of the BMN operators are
given again by (\ref{energ_parts_phase}) with the constant shift
of energies contributing  an extra  term $\ha \l\k_d^2M$
\begin{eqnarray}
E=\sum_{k=1}^M \epsilon_k=\frac{\l}{2L^2}\,\sum_{k=1}^M
\Bigg[\left(n_k +\beta_d L\right)^2 +  (\k_d L)^2 \Bigg]\=
\frac{1}{2}\,\lambda\,\sum_{k=1}^M \Bigg|\frac{n_k}{L} +\beta
\Bigg|^2\,.\la{enq}
\end{eqnarray}
This expression  indeed reproduces the leading order term in
the
expansion of  the energies of string states
in the plane wave limit of the deformed
background  which were derived in \cite{LUMA} (see also \cite{NIPR}).\footnote{
The operators dual to these string states can be
easily reconstructed using the details of the Bethe ansatz. It turns
out that the 2-impurity operators are identical to those in the
undeformed theory. }

Similarly to the case of real $\beta$ parameter,  the expression  \rf{enq}
can be obtained {}from the one in the undeformed  case  by shifting
 the mode numbers  by the (now complex)
deformation parameter $\beta$.
It is, however, clear that the real  and the imaginary  parts
of $\b=\b_d-i \k_d$ enter differently into the Bethe equations
and so the eigenvalues of the Hamiltonian depend on both the real and
imaginary parts of $n_k+\beta L$ but not only on $|n_k+\b L|$; this
 is  in a 
qualitative agreement with our conclusion
based on the string  world sheet analysis.


\

It is not difficult to obtain the integral Bethe equation
describing the thermodynamic limit of the spin chain for complex
$\b$. The same equation should be possible to derive {}from the
Landau-Lifshitz action for fast strings we found in \S 4.


\

\subsection{Comments on higher loop orders\label{HL}}

Let us now make few comments on higher-loop corrections to the
dilatation operator of the $\b$-deformed theory and possible
extension of the associated Bethe ansatz. Let us first recall that
in the $\b=0$ case  the 2-loop dilatation operator contains only
2-spin (nearest neighbour  and  next to nearest neighbour)
interactions \ci{BKS}
\be
\la{2loop}
\HH_2 = {\l^2 \ov (4 \pi)^4 }\sum^L_{l=1}
( -3 -  \bs_l \cdot \bs_{l+2}  + 4 \bs_l \cdot \bs_{l+1}) \ ,
\ee
and is expected to be part of an integrable
spin chain \ci{SEST,BDS}. It can be indeed embedded \ci{SEST} into
the integrable Inozemtsev \ci{INOZ} spin chain (with only 2-spin
interactions in the Hamiltonian) but the correct $\N=4$ SYM
dilatation operator in the $su(2)$ sector is different, containing
also 4-spin and higher interactions at three and higher loops
\ci{BKS,B}. The conjecture form of the corresponding  asymptotic
($L$ large compared to loop order) Bethe  ansatz was suggested in
\ci{BDS}
\begin{equation}
\label{bds}
e^{i p_k L}=\prod_{\stackrel{j=1}{j\neq k}}^{M}
\frac{\varphi_k-\varphi_j+i}{\varphi_k-\varphi_j-i}\;,
\end{equation}
were $p_k$ are the  magnon momenta  and the analogs of the rapidities
$\nu_k = \frac{1}{2}\cot\frac{p_k}{2}$ are  \ci{BDS}\foot{In the case of the Inozemtsev
chain
$\varphi(p) =\frac{1}{2}\cot\frac{p}{2}
+2\sum^\infty_{n=1}\frac{t^n \sin p}{(1-t^{n})^2+4t^n\sin^2 {p\ov 2} }$.
To match the 2-loop dilatation operator \ci{SEST}
  $t$ should be  related to the 't~Hooft coupling as
$
\frac{\lambda}{16\pi^2}=\sum^\infty_{n=1}\frac{t^n}{(1-t^n)^2}.$}
\be \la{dis}
\vp_k\equiv \vp(p_k)\ , \ \ \ \ \ \ \ \ \ \
 \varphi(p) = \frac{1}{2}\cot\frac{p}{2}\ \sqrt{ 1 + {\l \ov \pi^2}
\sin^2 { p\ov 2} } \ ,  \ee
with $E= \sum^M_{k=1} (  \sqrt{ 1 + {\l \ov \pi^2} \sin^2 { p_k\ov 2} } -1 ) $.

\

Turning now to the $\b$-deformed theory,
a  natural  series of questions are:
(i) Which is the structure of the  2-loop dilatation operator?
(ii) Is it integrable, e.g., is it possible to embed into some
integrable Inozemtsev-type spin chain with a Bethe ansatz,
whose solution, rewritten  as a series expansion in the
 't~Hooft coupling, reproduces both the 1-loop and
2-loop anomalous dimensions?\foot{Let us note  that {}from
a practical standpoint of the calculation of
anomalous dimensions, the  integrability of the 2-loop
dilatation operator
is not crucial, as one
can always treat the 2-loop corrections as a perturbation of the
1-loop integrable Hamiltonian; then
$\delta E = \langle E_{(1)} | \HH_{\rm 2-loop}|E_{(1)}\rangle
$,
where $|E_{(1)}\rangle$ stands for  an eigenstate of the
1-loop Hamiltonian with energy $E_{(1)}$.}
(iii) Is it possible to generalize the Bethe ansatz \rf{bds},\rf{dis}
to the case of non-zero deformation $\b$?

Starting  with the Feynman diagrams for the theory \rf{supe},\rf{vv}
one observes  that the
 2-loop dilatation operator  receives two types of contributions. The
first type corrects at order $\lambda^2$ the nearest neighbour
interactions already appearing at the 1-loop level in \rf{H}: they
 are given
by Feynman diagrams involving only two fields {}from the operator
whose renormalization we are studying. The second type involves
qualitatively new structures which can be interpreted as
 3-spin interaction
terms. They are  rather involved and we will not present them   here in
full generality.
Fortunately,   the important features   can
be extracted  already {}from only partial 2-loop  calculations.

\

Let us first discuss the pure
phase deformation, $\beta\in \Re$. In this case, an observation of
\cite{BECH} makes it easy to construct the  spin chain
Hamiltonian representing the 2-loop dilatation operator. First,
we note  that  by a position dependent phase transformation
the
1-loop Hamiltonian \rf{H}
can be mapped into the Hamiltonian of the XXX
Heisenberg chain, i.e. into the  1-loop  dilatation operator
in the $su(2)$ sector of ${\cal N}=4$ SYM.
This
can be formally implemented  as a  unitary transformation generated
by
\begin{eqnarray}
{\cal U}(\varphi)=
{\rm exp} \big[-\ha   i\varphi \sum_{k=1}^L\,k\,(1-\bs^z_k)\big] ~~,
\end{eqnarray}
which acts on the generators of $SU(2)$ as
\begin{eqnarray}
{\cal U}(\varphi)\bfsigma_l^\pm {\cal U}^\dagger(\varphi) = e^{\pm
i l\varphi}\bfsigma_l^\pm ~,  ~~~~~~~~~~~ {\cal
U}(\varphi)\bfsigma_l^z{\cal U}^\dagger(\varphi) = \bfsigma_l^z\ .
\end{eqnarray}
Second, in \cite{RR} it was noticed that the 4-scalar interaction in
\rf{vv}
coincides (up to  addition of the 
 identity operator) with the
1-loop Hamiltonian. As a result,
one  is able to argue that  the 2-loop spin
chain Hamiltonian for the deformed theory with $\beta\in\Re$ is the
same as the 2-loop spin chain Hamiltonian of the
${\cal N}=4$ SYM \rf{2loop},
 unitary-transformed with ${\cal U}^{-1}$.

Interpreting the Bethe equations as the cyclicity conditions for the
magnon excitations, it then follows that the effect of this ${\cal
U}^{-1}$
transformation should be  to {\it twist the boundary conditions}
 of the spin
chain. This suggests  the integrable extension of the
2-loop dilatation operator in the $\beta\in\Re$-deformed  ${\cal
N}=4$ SYM theory should be the same spin chain as in the absence of the
deformation but with twisted boundary conditions.
The corresponding asymptotic Bethe ansatz equations should then be
given by the following modification of \rf{bds}
\begin{equation}
\label{bdss}
e^{i (p_k- 2\pi \b_d)  L}= e^{-2\pi i \b_d  L} e^{i p_k  L}
  = \prod_{\stackrel{j=1}{j\neq k}}^{M}
\frac{\varphi_k-\varphi_j+i}{\varphi_k-\varphi_j-i}\  .
\end{equation}
In  this case the 1- and 2-loop matching  between   gauge and string theory
observed in undeformed case \ci{SEST,KRTS,KMMZ}  should extend
also to the real $\b$ deformation case.
Indeed, the subleading order $\l^2$ correction (\ref{2loopLL})
to the string reduced (Landau-Lifshitz) sigma model was found to depend
on $\b_d $ only through the combination $\eta' + \ha \b_d J$,
and thus   can be absorbed  by redefining
$\eta\to  \tilde \eta= \eta  + \ha \b_d J \sigma $, where the new field
satisfies  $
\tilde \eta (\tau, \sigma + 2 \pi) = \tilde \eta (\tau, \sigma ) +
\pi \b_d J $, i.e. has twisted boundary condition.

\

It is possible  that this ${\cal U}^{-1}$ transformation
extends to arbitrary number of loops and then  conjecture
that that the anomalous
dimensions of all operators in the 2-spin sector are given by the
the Bethe-like equations as in \cite{AFST} modified by the
inclusion of the phase $\exp(-2\pi i\beta_d)$ in the term
describing the expected phase shift $\exp{(i p_k L)}$ of a magnon
transported along the chain (see also the next section).

\

The case of general complex deformation parameter is by far
 less clear. One  very optimistic  guess  would be that  there exists an
integrable
spin chain like the  Inozemtsev chain
in which the exchange operator $P_{ij}= I - \bs_i \cdot \bs_j$
 is replaced by the 2-site
Hamiltonian in \rf{H} or (\ref{2site}).\footnote{In \cite{HALD} it was
conjectured that the Haldane-Shastry chain
(and then probably also the
Inozemtsev chain) has an anisotropic extension.}
It seems unlikely, however,
that the explicit calculations  will support this.
Indeed, looking at the Feynman diagrams,
a potential problem  is the appearance of 3-spin
interaction terms which cannot be put into  the form of a long range
2-spin interaction. Using the field-theory ingredients spelled out in
\cite{GMRO} and \rf{vv},\rf{vev} we can isolate such a term:
\begin{eqnarray}
(q_d^2-1)(\bfsigma^+\otimes\bfsigma^-+\bfsigma^-\otimes\bfsigma^+)
\otimes(\id-\bfsigma^z)\ .  \label{3spin_int}
\end{eqnarray}
The difference between the couplings of the fields $\Phi_i$
for zero and non-zero $\k_d$ prevents the term
with opposite signs of $\bfsigma^z$ {}from being generated and thus
this term seems to survive as a genuine 3-spin interaction.
The status of the integrability of such spin chains with
next-to-nearest neighbour interactions is not  clear.
Certain 3-spin deformations of the XXZ chain were  analyzed in
\cite{GRMA}, but none of the Hamiltonians considered there contain
(\ref{3spin_int}).
Even though no exhaustive search has been
carried out, we are tempted to expect  that the 2-loop spin chain
Hamiltonian of the
deformed  theory with complex $\b$ does not have an integrable extension.

There exists at least one  natural  reason to expect that the
integrability of the dilatation operator of the $\beta$-deformed
${\cal N}=4$ SYM is spoiled at some number of loops if
$Im(\beta)\ne 0$. The construction of the supergravity
dual of this type of deformation
required \ci{LUMA} the use of $S$-duality transformations.
Even though the resulting 
string coupling is small (due to the use of the {\it two}
S-duality transformations) this seems to  imply that
non-interacting strings in the deformed background  should
have a knowledge of interactions  in the undeformed theory.
We, however,  do not expect the dilatation operator
 of the $\N=4$ SYM
theory be integrable in such  finite $N$  regime.
The fact that the dilatation
operator in the 3-spin ($\Tr \Phi_1^{J_1} \Phi_2^{J_2} \Phi_3^{J_3}+...$)
 sector of the deformed theory is 
not integrable already at the 
1-loop level if $\sigma_d\ne 0$ may  be also 
an  indication  of this.



\renewcommand{\theequation}{7.\arabic{equation}}
 \setcounter{equation}{0}

\section{String Bethe Equations}\la{bethe}

In this section we shall use the Lax representation for strings on the
\LM background \rf{metrgen0}  with real $\tb\equiv\tg$ recently found in
\cite{SF} to derive the string Bethe equations in the
corresponding   ``$su(2)_\g$''
subsector of the model (in this section $\k_d=0$, i.e. $\b_d=\g$,
see \rf{kkk},\rf{ident3}.
 In this subsector we consider strings
moving on the deformed background $R\times S^3_\g$, described by
the action \rf{a1}.
By making the T-duality transformation on $\p_1$, a shift of
$\p_2$, followed by another T-duality on $\tilde \p_1$, we
get back to the standard action for a string
 on $R\times S^3$:
\bea \la{a2}
S = -{\sqrt{\lambda}\over 2}\int\, d\tau \int^{2\pi}_0  {d\s\over 2\pi}
 \g^{\a\b}\left[ -\pa_\a t\pa_\b t +\sum^2_{i=1} \left(
\partial_\a \r_i\partial_\b \r_i +
\r_i^2\partial_\a \ttp_i\partial_\b \ttp_i\right)\right]\, ,
\eea
where a double tilde over $\p_i$ reflects the two T-duality
transformations we performed.

It is not difficult to find the  following on-shell relations
between $\ttp_i$ and $\p_i$ \cite{SF} \bea \la{a3}
&&\partial_\a\ttp_1 =G\, \left(\partial_\a\p_1 -\hg\,\r_2^2
\g_{\a\b}\e^{\b\g}\partial_\g\p_2\right) \ ,\\ \nonumber
&&\partial_\a\ttp_2 =G\, \left(\partial_\a\p_2 +\hg\,\r_1^2
\g_{\a\b}\e^{\b\g}\partial_\g\p_1\right)\,. \eea By introducing
the momenta $p_i$  and $\tilde{\tilde{p_i}}$ conjugated to $\p_i$
and $\ttp_i$ respectively, the relations (\ref{a3}) can be
rewritten in the following simple form \bea \la{dphisu2}
\tilde{\tilde{p_i}} \= p_i\, ,\qquad \ttp_1^\prime  \= \p_1'\+
\g\,p_2\, ,\qquad \ttp_2' \= \p_2'\- \g\, p_1\, ,  \eea where $\g\=
{\hg\ov\sqrt{\l}}$ is the deformation parameter that appears on
field theory side. Taking into account that $\p_i$ are angle variables,
and integrating (\ref{dphisu2}) over $\s$, we get the following
twisted boundary conditions for the $U(1)$ variables $\ttp_i$ of
the $R\times S^3$ model: \bea \la{dphi2su2} &&\p_i(2\pi) \-
\p_i(0) \= 2\pi\, n_i\, ,\quad n_i
\mbox{ are integer winding numbers}\, , \\
\nonumber &&\ttp_1(2\pi) \- \ttp_1(0) \= 2\pi \left(n_1\+
\g\,J_2\right)\, ,\qquad \ttp_2(2\pi) \- \ttp_2(0) \= 2\pi
\left(n_2\- \g\,J_1\right)\, . \eea The relations (\ref{a3}),
(\ref{dphisu2}) and (\ref{dphi2su2}) imply that if $\p_i$ solve
equations of motion for strings on $R\times S^3_\g$ then $\ttp_i$
solve those on $R\times S^3$ with the twisted boundary conditions
(\ref{dphi2su2}) imposed on $\ttp_i$, and vice versa. One can also
check that if the Virasoro constraints are satisfied for strings
on $R\times S^3_\g$ then they are satisfied for twisted strings on
$R\times S^3$, and, therefore, the energy of a twisted string is
equal to the energy of the corresponding string on $R\times
S^3_\g$.

\vskip 0.3cm

Since the Lax pair for strings on $R\times S^3_\g$ is closely
related to the Lax pair for the sigma model on $S^3$, we recall
the necessary facts about the latter model. We follow closely the
discussion in \cite{KMMZ} to simplify the comparison of the string
Bethe equations for the string sigma model on $R\times S^3$ with
the equations we will derive for the $R\times S^3_\g$ model.

We parameterize  $S^3$  by  unitary $SU(2)$ matrices of the form:
\bea \la{matrsu2} g\=\left(
\begin{array}{cc}
X_1 & X_2 \\
-X_2^* & X_1^*
\end{array}
\right) \, , \qquad \det g \= |X_1|^2+|X_2|^2=1 \, ,\qquad X_i\=
\r_i e^{i\ttp_i}\, . \eea The equations of motion for the string
sigma model on $R\times S^3$ follow {}from the action (\ref{a2})
that can be cast in the form: \bea \nonumber {\rm
S}\={\sqrt{\l}\ov 4\pi}\int {\rm d}\tau{\rm d}\sigma \gamma^{\a\b}
\left[ \pa_\a t\pa_\b t + {1\ov 2} {\rm Tr}\Big(g^{-1}\,
\pa_{\a}g\,g^{-1}\, \pa_{\b}g \Big)\right] \, . \eea Introducing
the right current
$$
R_{\a}\=g^{-1}\,\pa_{\a}g\,  ,
$$
the equations of motion for the matrix field $g$ can be written in
the form \bea \label{eom} \pa_{\a}(\gamma^{\a\b}R_{\b})\= 0\, .
\eea They should be supplemented by the Virasoro constraints:
$$
\pa_0 t\pa_0 t + \pa_1 t\pa_1 t + {1\ov 2} \Tr\left(
R_0^2+R_1^2\right)=0\, ,\qquad \pa_0 t\pa_1 t + {1\ov 2} \Tr
(R_0R_1)=0\, .
$$
The equations (\ref{eom}) are equivalent to the zero curvature
condition \cite{lup,ZM,FR,FADT} \bea \label{zc} [D_{\a}\, ,\,
D_{\b}]\= 0 \, , \eea where the Lax operator depending on a
spectral parameter $x$ is defined as \bea \label{Lax}
D_{\a}\=\pa_{\a}\-\frac{R^{+}_{\a}}{2(x-1)}\+\frac{R^{-}_{\a}}{2(x+1)}\equiv
\pa_{\a}\- {\mathcal A}_{\a}(x) \, . \eea Here the self-dual  and
anti-self dual projections of $R_\a$ are given by \bea
\label{proj} R^{\pm}_{\a}\=(P^{\pm})_{\a}^{~\b}R_{\b},~~~~~~~~
(P^{\pm})_{\a}^{~\b}\=\delta_{\a}^{~\b}\mp
\gamma_{\a\rho}\epsilon^{\rho\b} \, . \eea
To obtain the Lax
representation for the $R \times S^3_\g$ model we perform the
following gauge transformation of the Lax connection
$\A_\a$\footnote{A similar idea was used in \cite{AF} to derive a
local and periodic Lax connection for the Hamiltonian of strings
on \adss.} \bea \la{gaugetrsu2}
&&D_\a\to \M\, D_\a\, \M^{-1} \, = \, \pa_\a \, -\, \RR_\a\, ,\\
\nonumber && \RR_{\a}\= \M\,{\mathcal A}_{\a}\,\M^{-1}\, -\,
\M\,\pa_{\a}\M^{-1}\, =\, \hat{{\mathcal A}}_{\a}\, +\, {i\ov
2}(\pa_{\a}\ttp_2 -\pa_\a\ttp_1)\, \s_3\, , \eea where \bea \la{M}
&&\M\ = \left(\begin{array}{cc}
0 & e^{{i\ov 2}(\ttp_2 -\ttp_1)} \\
e^{-{i\ov 2}(\ttp_2-\ttp_1)} & 0
\end{array}
\right)\, ,\quad \ \ \ \ \M^{-1}\= \M\, . \eea One can show that the new Lax
connection $\RR_\a$ depends on $\ttp_i$ only through
$\pa_\a\ttp_i$ \cite{SF}. The local and periodic Lax connection
for the $R \times S^3_\g$ model is now obtained by expressing the
twisted angle variables $\ttp_i$ of $S^3$ in terms of the angle
variables $\p_i$ of $R \times S^3_\g$ by using the relations
(\ref{a3}).

\vskip 0.3cm

To derive the string Bethe equations we need to analyze various
asymptotic properties of the monodromy matrix T$(x)$ which is
defined as the path-ordered exponential of the spatial   component
$\RR_{1}(x)$ of the Lax connection
\bea \la{T} {\rm T}(x)={\cal P}\exp\int_0^{2\pi}{\rm
d}\s~ \RR_{1}(x) \, , \eea and also of
the quasi-momentum $p(x)$ defined by
\bea
\la{quasi} 2\cos p(x) = \Tr\, {\rm T}(x)\ . \eea It is well-known
that if the Lax connection is periodic then the quasi-momentum
$p(x)$ is conserved for any value of the spectral parameter $x$,
and it plays an important role in the inverse scattering method
\cite{RS}.

The Lax connection has  poles at $x\= \pm 1$, and the  analysis
of the asymptotic behavior of the quasi-momentum around the poles
does not,  in fact,  require any computation. All one needs to do is
to notice that around the poles one can diagonalize the Lax
connection $\RR_1$ by means of a regular gauge transformation
which depends only on $\pa_\a\ttp_i$, and, therefore, is periodic.
This gauge transformation diagonalizes the   Lax connections both
for the $R\times S^3_\g$ and the $R\times S^3$ models. Therefore, the
asymptotic behavior of the quasi-momentum in  both models is
the same \cite{KMMZ} \bea \la{asympm1} p(x) \= - {\pi \k\ov x\pm
1} +\dots  \qquad  x \to \mp 1\, , \eea where $\kappa$ is the
rescaled energy of the string solution \bea \la{kapp} E =
\sqrt{\l}\k\, , \qquad \k = -\int_0^{2\pi}{{\rm d}\s\ov 2\pi}\,
\g^{0\a}\pa_\a t\, . \eea In what follows, we set,  without loss of
generality,  $\g^{\a\b} = \mbox{diag}(-1,1)$ and $t = \k
\tau$.

The analysis of the
asymptotic behavior of the quasi-momentum at infinity and zero is more involved
because it does not vanish at large values of the spectral
parameter $x$, and we present its details in the Appendix B. The results
of this  analysis are summarized below:
\bea
\la{largex}
p(x)&\=&  \pi\g L \- {2\pi(L-2M)\ov
\sqrt{\l} x} \+ \dots \, ,\quad x\to \infty \, ,\\
\la{smallx}
p(x)&\=& \pi\g (L-2M) \+ {2\pi L\ov \sqrt{\l} }\, x \- 2\pi m\+ \dots \,
,\quad x\to 0 \, ,
\eea
where
$$ L \equiv J= J_1 + J_2 \ , \ \ \ \ \ \ \
M\equiv J_2  \ , $$
and  $m$ is an integer quasi-momentum number.

Subtracting the singularities at $x\to \pm 1$, and the constant
term at $x\to\infty$, we define the resolvent \bea \la{resolv}
G(x)\= p(x) \+ {\pi\k\ov x-1}\+ {\pi\k\ov x+1}\- \pi\g L\, . \eea
Following \cite{KMMZ}, we assume that the resolvent is an
analytical function on the complex plane of the spectral parameter
$x$ with a finite number of cuts $C_i$. Then, it admits the
following spectral representation \bea \la{resolv2} G(x)\= \int_C
{\rm d}\xi\, {\r(\xi)\ov x-\xi}\, , \qquad C = C_1\cup C_2\dots
\cup C_n\ \,, \eea where $\r(\xi)$ is a positive spectral density.

The resolvent has the following asymptotic behavior at infinity
and zero \bea \la{resolv_asym} &&G(x) \= {2\pi \ov \sqrt{\l}\, x}
(E + 2M- L) \+\dots\, ,\qquad x\to\infty\\ \nonumber &&G(x) \=
-2\pi\g M \-2\pi m \-{2\pi x \ov \sqrt{\l}} (E - L) \+\dots\,
,\qquad x\to 0\, . \eea This asymptotic behavior  leads to the
following constraints imposed on the density
\bea \la{dens1}
\int_C {\rm d}\xi\, \r(\xi)&\=&{2\pi \ov \sqrt{\l}} (E + 2M-
L)\,,\\
\la{dens2}
 -{1\ov 2\pi i}\oint{G(x)\ov x}{\rm d}x\=\int_C {\rm d}\xi\,
{\r(\xi)\ov \xi}
&\=& 2\pi\g M\+2\pi m\,,\\\la{dens3}
-{1\ov 2\pi i}\oint{G(x)\ov
x^2}{\rm d}x\=\int_C {\rm d}\xi\, {\r(\xi)\ov \xi^2} &\=& {2\pi
\ov \sqrt{\l}} (E - L)\,.
\eea
Finally, the quasi-momentum obeys
the unimodularity condition \cite{KMMZ}
\bea \la{unimod} p(x+i0)\+
p(x-i0)\= \-2\pi n_k\,,\qquad x\in C_k\,,
\eea
that leads to the following singular integral equation for the
spectral density
\bea
&& G(x+i0)\+ G(x-i0) \nonumber \\
&&= \ 2\, \pint_C {\rm d}\xi\,
{\r(\xi)\ov x-\xi}\= {2\pi\k\ov x-1}\+ {2\pi\k\ov x+1}\- 2\pi\g L
\- 2\pi n_k\,,\ \ \ \quad x\in C_k\,.
\label{inteq}
\eea

Comparing equations (\ref{dens1}), (\ref{dens2}), (\ref{dens3})
and (\ref{inteq}) with those for the undeformed case \cite{KMMZ},
we see that they differ only by the  $\g$-dependent shifts of the
quasi-momentum number $m$ and the mode numbers $n_k$.
 Therefore,
the comparison of these equations with the thermodynamic Bethe
equations for the $\g$-deformed spin chain  we derived in section
\ref{spin_chain_stuff}  repeats the comparison done in \cite{KMMZ}
for the undeformed case.\footnote{To perform the comparison one
should first rescale $x$ and $\xi$: $ (x,\, \xi) \to {4\pi L\ov
\sqrt{\l}}\, ( x,\, \xi)\, , $}
We conclude that the string
Bethe equations coincide  with the
thermodynamic Bethe equations for the $\g$-deformed spin chain
at the  one-loop ($\l$)  and two-loop ($\l^2$) orders.
This constitutes
a proof of  two-loop  agreement between  the string and the
gauge theory results in the $su(2)_\g$ subsector.

Moreover, it is straightforward to write down the quantum (i.e.
discrete) version
of the classical Bethe equations following  \cite{AFST},
and assuming that the phase function appearing in the Bethe ansatz
for the $\b$-deformed spin chain coincides with the phase function
of the asymptotic Bethe ansatz for the $su(2)$ subsector of $\N =
4$ SYM \cite{BDS}.\footnote{The phase function was derived in
\cite{BDS} assuming the validity of the BMN scaling.
Let us note that a  recent
computation \cite{Plefka} of the 4-loop dilatation operator in a matrix model
related to $\N=4 $ SYM by dimensional reduction on $S^3$
suggests  that the BMN scaling may be broken
(in the matrix model) starting at 4-loop order.}

 It is natural to expect that, as it was the case in the
undeformed case \cite{AFST}, the Bethe equations for quantum
strings will  reproduce the $1/J$ correction to the BMN states.
It would be interesting to check
this by a direct string theory computation generalizing the one
done in \cite{Callan}. The string theory action  can be obtained
(up to the quartic
order in string fields including fermions)
{}from the
string action on $AdS_5\times S^5$ by using the
TsT transformation  we discussed in sections 1 and 2.
It would be also interesting to understand how the
relation between the $su(2)$, $sl(2)$ and $su(1|1)$ subsectors of
$\N =4$ SYM found in \cite{Staud} is modified in the
deformed case.


\renewcommand{\theequation}{8.\arabic{equation}}
 \setcounter{equation}{0}

\section{ Subleading $1/J$ corrections
\label{subleading}}

Given the success in  the comparison of  the 1-loop
spin chain and fast string expansion of the sigma model,
it is natural to ask whether it can be extended to subleading terms
in the large quantum number expansion. Then on the gauge theory / spin
chain side we are interested in finite size corrections to the
thermodynamic limit of the Bethe ansatz while on the string theory
side we are interested in world sheet quantum corrections
 \ci{FT,FRPATS,BTZ}.

An interesting fact which emerged from the calculation of
the leading $1/J$ correction term in the  world sheet
1-loop correction to the energies of semiclassical circular string solutions in
$AdS_5\times S^5$ \cite{FRPATS} is that they are reproduced by 
the 1-loop
calculation in the Landau-Lifshitz theory,  provided one uses the 
$\zeta$-function  regularization to compute the infinite  sum
over characteristic frequencies 
\cite{BTZ}.\foot{The reason for that is that the relevant contribution
comes essentially  from the characteristic frequencies in the
directions included in the Landau-Lifshitz action, while
the role of  other fluctuations in other bosonic and fermionic directions
is to regularize the resulting infinite sum.}

In the case of the $\beta$-deformed theory,
the analysis of the 1-loop corrections to the string
energies, while certainly doable,
 is somewhat complicated by the large number of fluxes
in the supergravity  background.  In the following we will not
perform  the full
string theory calculation; rather, we will follow the 
example of  the undeformed theory 
and use the ``fast string''
expansion of the string sigma model. 
We will compute the 1-loop correction
to the energy of circular 2-spin solution
in this theory using the $\zeta$-function regularization and compare
it with the leading finite size correction to the thermodynamic limit of
the Bethe Ansatz. We will find that, as in the case of the undeformed
theory \ci{BTZ}, the two  expressions
 are in perfect agreement.

\subsection{String sigma model  side}

In general, computing  the quantum shift in the energy of some classical solution
requires finding and summing up the frequencies of the fluctuations
around it. The ``fast string'' limit of the sigma model has some
simple classical solutions for which it is possible to do this
analysis very explicitly.

The case of real deformation parameter is again very easy to analyze.
As we have seen  in section  4.2, the effect of
such  deformation is to twist the boundary conditions of
the $\eta$ field  in the world sheet $\sigma$ direction (see \rf{LL}).
We, therefore,  expect that in this case the frequencies of the
fluctuation modes near the circular solution of section 4.1 or \rf{new}
 should follow  {}from the results of \cite{BTZ}
for $\b=0$ theory  by
shifting the winding number $m$ by $\ha \beta_d J$
\begin{equation}
 E^{(1)}_1 =
{\lambda \ov 2J^2}
{ (m+\ha \beta_d J)^2} + { \lambda \ov 2 J^2}
 \sum^{\infty}_{n=1}
 \left[
n \sqrt{ n^2 - 4(m+\ha \beta_dJ)^2}
-   n^2 + 2 (m +\ha \beta_d J)^2
\right]~~.
\label{origshift2}
\end{equation}
This expectation is, of course, realized. We will recover the
equations above as a limit of the frequencies in the presence of a
complex deformation.

The Lagrangian  \rf{LL} or \rf{wz},\rf{kop} expressed  in terms of the  unit vector
${\vec n}(\theta,\eta)$ in \rf{coh_st} is
\begin{eqnarray}
\label{LL_ac}
L&=&L_{WZ}
-\frac{\tilde\lambda}{8}
\left[
{\bf n}'\,{}^2
-\big|\beta J\big|^2
(n_z^2-1)
+{2\beta_dJ}(n_x n_y'-n_y n_x')
\right]~~.
\end{eqnarray}
Using that
$\delta L_{WZ}=\frac{1}{2}\epsilon_{ijk}\delta n_in_j{\dot n}_k$
leads to the equations of motion
\begin{equation}
\frac{1}{2}\epsilon_{ijk}n_j{\dot n}_k=
-\frac{\tilde\lambda}{4}
\left(
n_i''+v_i
\right)_\perp
~~\longrightarrow~~
{\dot n}_i=
\frac{\tilde\lambda}{2}
\epsilon_{ijk}n_j\left(
n_k''+v_k
\right)
\label{LL_eq}
\end{equation}
where the vector $v$ is given by
\begin{eqnarray}
{\bf v}=
J\,(-{2\beta_d}\,n_y',2{\beta_d}\,n_x',
J{|\beta |^2}n_z)~~.
\end{eqnarray}
and $\perp$ denotes the projection orthogonal to ${\bf n}$.
When  the deformation is removed,  ${\bf v}$ vanishes and the equations
(\ref{LL_eq})
become the standard isotropic Landau-Lifshitz equations.

We have seen  in  sections 4 and 5 that the terms linear in $\sigma$
derivatives are producing the  twisting of the boundary conditions of the
world sheet field $\eta$. Taking this standpoint (that is, removing
the first two components of ${\bf v}$ and recalling that the winding
number should then be shifted by $\ha J\beta_d$) we see that the
equations above are a special case of the anisotropic
Landau-Lifshitz equations, which are known to be integrable
\cite{FADT}. As was already noted above,
this  squares nicely with the Lagrangian  (\ref{LL_ac}) being
the continuum limit of the coherent state action for
the  integrable spin chain.

Let us now consider fluctuations near the solution \rf{new} with
$J_1=J_2$. 
This is straightforward to do in terms of the two angular coordinates,
but the result may be presented also in terms of $\bf n$ parametrization.
If we  consider the solution (cf. \rf{new},\rf{coh_st})
\begin{eqnarray}
{\bf n}=(\cos 2m\sigma,\,\sin 2m\sigma,\, 0)~~,
\label{states}
\end{eqnarray}
then parameterizing the fluctuations as
\begin{equation}
\delta {\bf n}=(-\sin 2m\sigma\,A_1(\tau,\sigma),\,
\cos 2m \sigma\,A_1(\tau,\sigma),\, A_2(\tau, \sigma))~~,
\end{equation}
it is a simple exercise to  linearize the equations of motion.
All world sheet position dependence disappears and we are left with
two coupled Schr\"odinger-type equations
\begin{equation}
{\dot A}_1=\frac{\tilde\lambda}{2}
~\left[
A''_2+4\big|m + \ha \b J \big|^2A_2
\right]\ , \ \ \ \ 
~~~~~~~~ {\dot A}_2=\frac{\tilde\lambda}{2} ~A''_1~~.
\end{equation}
Introducing the mode expansion
$A_{s}\propto\sum C_{s,n}e^{i\omega_n\tau +
in\sigma }$ (the boundary conditions on the fluctuations are not
twisted in this parametrization) we find the mode frequencies:
\begin{equation}
\omega_n
=\frac{\tilde\lambda}{2}
~
n\sqrt{n^2-4\Big|m +\ha { \beta}{J}\Big|^2}\ .
\label{freq}
\end{equation}
This result is very similar to the one in the absence of the
deformation.\foot{It is interesting to note that
while the initial field equations depended
separately on the real and imaginary parts
 of the deformation $\b$,
 the frequencies of the
fluctuations happen to depend only on the modulus of complex  $\beta$
shifted by $2m/J$. Most  likely this is  not a general property
but rather a special feature of the simple solution \rf{new}.}
The energy shift is given by the sum of the frequencies (\ref{freq}).
Applying then the $\zeta$-function regularization as in \cite{BTZ},
we conclude that the 1-loop correction to the energy of the classical string
solution (\ref{states})
is given by the same expressions as in \cite{BTZ}
with  the winding number $m$ replaced by by
$|m+\textstyle{\frac{1}{2}}{\tilde \beta}{\cal J}\Big|$
\begin{equation}
 E^{(1)}_1 =   E_{\rm reg}   + E_{\rm fin}~~,~~~~~~~~
E_{\rm reg} =
{ \lambda \ov 2 J^2}
 \left[\sum^{\infty}_{n=1}  ( n^2 -
2\Big|m+\ha {\beta}{J}\Big|^2 )
\right]_{\rm reg}
=\frac{\lambda}{2J^2}
{\Big|m+\ha {\beta}{J}\Big|^2}
\end{equation}
\begin{equation}
 E_{\rm fin}= { \lambda \ov 2 J^2}
 \sum^{\infty}_{n=1}
 \bigg( n \sqrt{ n^2 - 4
\Big|m+\ha {\beta}{J}\Big|^2}
-   n^2 + 2 \Big|m+\ha {\beta}{J}\Big|^2 \bigg)\ .
\label{energ_cpx}
\end{equation}
It is gratifying to see that in the $\sigma_d\rightarrow 0$ limit this result
matches our expectation (\ref{origshift2}).

According to the discussion in \S\ref{limpar} these corrections should
be compared to the finite size contributions to the energy of spin
chain eigenstates. We now discuss their calculation following the same
steps as in the undeformed case \ci{BTZ}.


\subsection{Spin chain side}

As should be obvious 
 by now, it should be  quite easy to find 
the $1/J$
corrections to the energies of spin chain states 
in the presence of a
real deformation parameter.
 This is   clear {}from the discussion of the Bethe ansatz equations
(see
(\ref{tw_Bethe_cyclic}) and \S\ref{bethe})
that the structure of the  subleading terms in the $1/J$ expansion is  not
affected by the shifts of  the mode number  by $\beta$.
Thus, the
agreement between the string and gauge theory results
 found in \cite{BTZ} should go over to the real $\b$ deformation
 case without any modification.

For  complex $\beta$ the situation is less clear and
we will consider it in some detail.
The conclusion of the
previous subsection was that the 1-loop  correction to the
classical energy of the string solution (\ref{states}) depends on the
deformation parameter $\beta$ only in the combination
$|m+\frac{1}{2}\beta J|$.
At the same time, the  Bethe equations (\ref{BE}),
(\ref{energ_parts}) depend separately  on the real and imaginary
parts of the deformation parameter;  it seems 
that a small miracle
is needed for them to combine to produce
the energy as a  function of the complex $\beta$ as in \rf{energ_cpx}.
Remarkably, this is indeed what  happens.

As in section 6, in  this subsection we shall use the notation
$L\equiv J=J_1+J_2, \  M= J_2$; the $J_1=J_2$ 
case of the previous subsection corresponds to $L=2M$.
The starting point of the calculation is the thermodynamic limit
 of the
Bethe equations and the cyclicity condition (\ref{BE}).
The leading
contribution can be extracted without much complication
(\ref{BE_large_L}). To properly
take into account the configuration of roots with spacing of
order  $1/L$ and find the leading finite size correction to the
energy we follow \cite{BTZ} and use an integral representation for
the logarithm of the factors in (\ref{BE}) depending on the difference
of rapidities. Focusing on the solutions with equal mode numbers
for 
all $M$ roots
($n_1=n_2\dots=n_M$ in (\ref{sh_T_lim})),
 the Bethe equations then become
\begin{eqnarray}
-{2\pi(n_1+\beta_d L)} +\frac{1}{x_k}
=2
\sum_{j\ne k=1}^M\int_{0}^{1/L}
\frac{f(x_k,x_j)\,\, d\epsilon}
{\epsilon^2 +(f(x_k,x_j))^2}\ , 
\label{tbl_subl}
\end{eqnarray}
where on the left hand side we kept the first two nontrivial 
orders in
the $1/L$ expansion and introduced  
the function $f(x_k,x_j)$ 
\begin{eqnarray}
f(x_k,x_j)=\frac{x_k-x_j}{1+\K^2x_kx_j}~~, \ \ \ \ \ \ \ \ \ 
\K\equiv  2\pi\kappa_dL\ , 
\end{eqnarray}
where $\K$
 is kept  finite in the limit  
$L\rightarrow\infty$. 
{}From  \rf{tbl_subl} it is  clear that the change of variables
(\ref{change_nu}) 
isolates  the difference  between the deformed (XXZ chain) 
and undeformed
(XXX chain) cases 
and making  it easy to compare the two calculations.

Once  $x_k-x_j$ is  of  order of $1/L$
the long-chain expansion becomes unreliable as the two terms in the
denominator become of the same order
 and the integral gives an ${\cal
O}(1)$ contribution to (\ref{tbl_subl}). To take into account 
such 
configurations we define  the resolvent
\begin{eqnarray}
G(x)=\frac{1}{L}\sum_{k=1}^M\frac{1}{f(x,x_k)}=\frac{1}{L}
\sum_{k=1}^M\frac{1+\K^2 x x_k}{x-x_k}\rightarrow \int d\xi
\,\rho(\xi)\,
\frac{1+\K^2x\xi}{x-\xi}
\label{res_init}
\end{eqnarray}
where we introduced the density of roots $\rho(\xi)$. The density
satisfies the condition that $\rho(\xi)d\xi$ is real and positive
and normalized to the filling fraction $\alpha=M/L$.
As in the case of the XXX chain \cite{BTZ}, our goal will be to
rewrite (\ref{tbl_subl}) in terms of the resolvent.
Direct comparison with the
calculation of \cite{BTZ} 
can be done at every step by taking the limit
of vanishing deformation. Indeed, as $\K\rightarrow 0$
(\ref{res_init}) reduces to the resolvent used in \cite{BTZ}.

It is important to notice that the resolvent \rf{res_init}
we introduced here does
not vanish at infinity.  Thus, it is not the density of roots but
rather a rescaled density of roots which is related to its
discontinuity. The expression 
for  the energy in terms of the density
is changed  compared to (\ref{dens3}), and this change  
 can be
compactly taken into account by expressing the energy directly in
terms of $G$. Then the energy of the spin chain 
states accessible
in the thermodynamic limit and the cyclicity constraints
\rf{BE},(\ref{energ_parts}) become:
\begin{eqnarray}
\label{en_total}
E&=&\frac{\lambda}{8\pi^2}\frac{1}{L^2}\sum^M_{k=1}
\frac{1+\K^2x_k^2}{ x_k^2}
=-\frac{\lambda}{8\pi^2 L}\,
\,G'(0)\ , \\
G(0)&=&-{2\pi(m+\beta_dM)} =
-{2\pi(n_1+\beta_dL)} \,\frac{M}{L}~~.
\label{G0}
\end{eqnarray}
The first expression for  $G(0)$ is obtained from 
 the thermodynamic limit
of (\ref{BE}) while the second form 
is obtained by summing up the  $k$
equations (\ref{tbl_subl}). Thus (cf. \rf{mem})
\be    m L = n_1 M   \ . \la{mmm} \ee
In particular, for the $J_1=J_2$ case ($L=2M$) we have 
$m= \ha n_1$. As discussed in \ci{BTZ}, 
while on spin chain side one may consider states with arbitrary
$ m=n_1 M/L$, only states with integer $m$ correspond to 
operators of $\N=4$ SYM and thus also string-theory states.

The goal of rewriting (\ref{tbl_subl}) in terms of $G(x)$ as well as of
making a consistent $1/L$ expansion can be achieved by multiplying
(\ref{tbl_subl}) by the generic term in the resolvent and summing over
$k$.
After a few simple
algebraic manipulations, we find that the content of the
equations (\ref{tbl_subl}) is captured by
\begin{eqnarray}
&&-{2\pi(n_1+\beta_d L)}G(x)
+\K^2 \frac{M}{L}\left(1-\frac{M}{L}\right)
-G^2(x)+\frac{1}{x}\Big(G(x)-G(0)\Big)\nonumber \\
&&=
\frac{1}{L}(1+\K^2x^2)\bigg[
{G'(x)}
-
\sum_{\stackrel{j\ne k}{j, k=1}}^M
\frac{1+\K x_kx_j}{(x-x_k)(x-x_j)}
\int_0^{1/L}\frac{\epsilon^2~~~d\epsilon}
{\epsilon^2  + (f(x_k,x_j))^2}
\bigg]~. \label{be_L}
\end{eqnarray}
This equation cleanly separates the terms contributing to finite size
corrections from those surviving in the infinite chain limit for any
configuration of roots. Indeed, the right hand side is manifestly of
order $1/L$.  This allows us to find the resolvent perturbatively in
$1/L$ by making the ansatz
\begin{eqnarray}
G=G_0+\frac{1}{L}G_1+\dots ~~.
\label{pert_exp}
\end{eqnarray}
 Imposing the correct behaviour (\ref{G0})
at the origin, the leading order resolvent $G_0(x)$ is given by
\begin{eqnarray}
G_0(x)&=&\frac{1}{2}\Bigg[\frac{1}{x}-2\pi(n_1+\beta_dL)\nonumber \\
&-&\frac{1}{x}\sqrt{
\left[1 - 2\pi(1-2\alpha)(n_1+\beta_d L)\,x\,\right]^2 +
4\,\alpha(1-\alpha)\Big|2\pi (n_1+\beta L)\Big|^2\,x^2
}\ \Bigg] ~,
 \label{loG}
\end{eqnarray}
where 
$$ \a \equiv{ M\ov L} \ . $$
Using (\ref{en_total}) it is then a simple exercise to derive the
energy of this spin configuration
\begin{equation}\la{eeee}
E_0=-\frac{\lambda}{8\pi^2 L}\,G'_0 (0) =
\frac{\lambda}{2 L} \alpha(1-\alpha)\Big|n_1+\beta L\Big|^2~~.
\end{equation}
This is the same expression we found for the corresponding 
Landau-Lifshitz solution \rf{en_gen} (where $\mm=n_1$, $L=J$). 
{} Note that while on  the ``fast string'' action side  \rf{LL}
it is clear that
the energy should depend only on $\beta$,  the initial
form of the Bethe equations obscures this property, 
 which is
restored only at the end of the calculation.

Let us now analyze the subleading corrections. The correction $G_1$
to the leading order resolvent is easily found by plugging
(\ref{pert_exp}) into the left hand side of (\ref{be_L}) and keeping
only the terms of  order  $1/L$. In spite of the form of
(\ref{loG}), $G'_1(0)$  has a simple expression
\begin{equation}
{G'_1(0)}={G_0'(0)}
-
\sum_{\stackrel{j\ne k}{j, k=1}}^M
\frac{1+\K^2 x_jx_k}{ x_j  x_k}
\int_0^{1/L}\frac{\epsilon^2~~~d\epsilon}{\epsilon^2
+ (f(x_k,x_j))^2} \ . 
\label{subl_res}
\end{equation}
To do the two summations we first notice that since we need only the
leading order contribution of the integral, the only
relevant rapidity configurations are those in which $x_j-x_k$
is of order $\epsilon\sim 1/L$. As in \cite{BTZ} we will approximate
this difference by a linear function weighted by the leading-order
density of roots
\begin{equation}
x_j-x_k\simeq \frac{j-k}{L\rho_0(x_k)}~ \ , 
\label{approx}
\end{equation}
and replace $x_k$ with $x_j$ elsewhere. It is then easy to do the
summation over the difference $(j-k)$. In the process,  the term
$G_0'(0)$ in \rf{subl_res} cancels out and  we are left with
\begin{eqnarray}
{G'_1(0)}= -
\sum_{ k=1}^M
\frac{1+\K^2x_k^2}{x_k^2}
\int_0^{1/L}\,d\epsilon
\left[\pi L (1+\K^2x_k^2)\rho_0(x_k){\epsilon}\right]
\coth\left[\pi L (1+\K^2x_k^2)\rho_0(x_k){\epsilon}\right]~,
\end{eqnarray}
where the factors of $(1+\K^2x_k^2)$ under the integral sign come from
the denominator of $f(x_j,x_j)$ in (\ref{subl_res}). Changing the
integration variable  to $\xi=\pi L(1+\K^2x_k^2)\rho_0(x_k)\epsilon$
and converting the remaining sum into an integral using the measure
defined
by (\ref{approx}), i.e.  $\sum_k 1/(L\rho(x_k))\rightarrow \int dy$,
 we end up with
\begin{eqnarray}
{G'_1(0)}=-\frac{1}{\pi}\int
\frac{dy}{y^2} \int_0^{\pi (1+\K^2y^2) \rho_0(y) }\,d\xi
\,\ \xi\,\ \coth\xi\ , 
\label{final_0}
\end{eqnarray}
where the $y$-integral runs over the support
 of the density function.

As already mentioned above, 
the  density of roots is usually equal to  the discontinuity of the
resolvent across its cut, but with 
 our definition of the resolvent
(\ref{res_init}) it is  $(1+\K^2y^2) \rho_0(y)$ that has
this property.
$G_0(x)$ found in (\ref{loG}) does not have a cut on
the real axis, but rather for complex rapidities. Thus, as in
 the case
of the rational solution of the  XXX chain 
 case (i.e. the $su(2)$ sector of ${\cal N}=4$ SYM),  
 the roots of the Bethe
equations do not lie on the real axis. The endpoints of the cut are
determined by zeros of the square root in $G_0(x)$ (\ref{loG}), and
the cut  intersects the real axis at some point.
Then the density function is
\begin{eqnarray}
(1+\K^2y^2)\rho_0(y)&=&\frac{1}{2\pi i}\Big(G_0(y+i 0)-G_0(y-i
0)\Big)\cr
&=&
\frac{1}{2\pi i\,y}\sqrt{(4 \pi \M\,y)^2 +  \left(1- z_0 \, y\right)^2}\ , 
\label{density}
\end{eqnarray}
where 
\be \la{MMM}
\MM \equiv  \sqrt{\alpha(1-\alpha)} \  |(n_1+\beta L)\Big| \ , \ee
and  $z_0$  is related to the intersection point
between the support of the root density (the cut) and the real 
(its 
value will  not be important for us). This density is
the same as in the case of the undeformed theory, except for 
the terms
proportional to $\beta$ under the square root which enter only 
through 
$|n_1+\beta L|$. There is no additional dependence on the
real or imaginary parts of $\beta$ in $G'_1(0)$ and in the
energy. Thus it is clear that the finite size
corrections we find can be obtained from 
those of the XXX chain by
replacing the mode  number $n_1$  by $|n_1+\beta L|$. This
is exactly the same as the result of the analysis of 
the fluctuations
of the fast string action in the previous subsection.

Changing the integration variable
 to $y'=1/y-z_0$ we bring the endpoints of the support of
the root density to  the imaginary axis. Since the integrand in
(\ref{final_0}) is an analytic function we can
deform the cut so that to
place it on the imaginary axis. Then, making explicit 
that the new integration variable is purely imaginary,  $y'=i z$,
 and
also redefining $\xi=-i\zeta$, the integral in \rf{final_0} 
becomes
\begin{eqnarray}
{G'_1(0)}=-\frac{1}{\pi}\int_{z_-}^{z_+}
{dz} \int_0^{\frac{1}{2}
\sqrt{(4 \pi \MM)^2 -z^2 } }
\,d\zeta \,\ \zeta\,\ \cot\zeta\ , 
\label{final_1}
\end{eqnarray}
where $z_\pm$ are the 
endpoints of the support of the density function 
in
these coordinates, $z_\pm=\pm 4 \pi \MM$.
Trading the integration over $z$ for an integral over the
upper limit of the $\zeta$ integral and 
 integrating  by parts
we finish  with the following 
expression for  the derivative of the first finite size
 correction to the resolvent 
%
and thus  the energy shift in \rf{en_total} 
\begin{eqnarray}
\label{fin_size}
E_{1}
=-\frac{\lambda}{8\pi^2 L^2}\,G_1'(0)
=\frac{2\lambda}{L^2}\,  \MM^{3}
\int_{-1}^1 dx\,\ x\ 
\sqrt{1-x^2 }\ \cot\left(2\pi\MM x\right)\ . 
\nonumber
\end{eqnarray}
This is the same expression as in \ci{BTZ} with $n_1 \to 
| n_1 + \b L|$ in $\MM$  defined in \rf{MMM}.

Representing  the cotangent as $\pi a \cot (\pi a) =-
\sum^{\infty}_{n=-\infty} {a^2\ov n^2-a^2}$
and integrating the resulting terms we get the same series 
expression for the correction to the energy as in 
 (\ref{energ_cpx}) where  $m= \ha n_1$ and $L= 2 M = J$.


\renewcommand{\theequation}{9.\arabic{equation}}
 \setcounter{equation}{0}

\section{Summary and further directions}\label{con}

In this paper we have shown that the agreement between the gauge
theory and the string theory calculations of the anomalous dimensions of
``long'' 2-scalar operators in ${\cal N}=4$ SYM survives certain exactly
marginal deformations of the gauge theory.
The supergravity background dual to the gauge theory with a real
deformation parameter (as defined in (\ref{supe})
and (\ref{kkk})) was constructed in \cite{LUMA} using a combination of
T-duality transformations and an isometric shift. We have argued that
the world sheet Green-Schwarz model in this background is integrable.

The relation between this background and $AdS_5\times S^5$
also implies that classical
solutions in the deformed and undeformed theories
are mapped onto each other;
 as an example, we have presented 
  the
deformed 
 2-spin ``circular'' string solution which is direct counterpart of 
the  one  in \cite{FT}.
We have shown that after properly 
identifying the parameters on the
two sides of the duality, it is possible to match the ``fast string''
expansion in the deformed
background (the analog of the Landau-Lifshitz action) 
  with the continuum limit of the 
  coherent-state action for the corresponding  spin chain.
 Also, 
 the gauge theory spin chain correctly reproduces the ``small
string'' limit of the sigma model (the analog of the BMN limit
in $AdS_5\times S^5$).
We have also speculated about  the 2-loop structure of the dilatation
operator in the 2-spin  sector.
Starting from the sigma model on $\Re\times S^3_\beta$
we have derived the string Bethe equations for real deformation
parameter and found that they are remarkably close to those in the
undeformed background. Finally, we have also analyzed 
 ``finite size''
corrections to the anomalous dimensions of  
operators dual to circular
rotating  strings and
found that they continue
 to reproduce the world sheet predictions even in
the presence of the deformation.

In a nutshell, the correspondence  between the gauge theory and
the string theory in this deformed, less supersymmetric,  but 
still conformal case 
 works very well -- perhaps better than we had the 
right to expect. 
Our analysis has a number
 of interesting extensions discussed below.

In the case of the $AdS_5\times S^5$ -- ${\cal N}=4$ SYM duality an
underlying reason for this ``semiclassical'' agreement is
 the 
 equivalence of
the two integrable models governing the leading order
corrections. For a real deformation parameter ($|{\bf q}|=1$ in the
notation of section \ref{rev}) we have seen that the same is still
true. In particular, 
this  suggests that there exist
many more sigma models which are classically integrable even
 though
their target space is not a coset space. 
After all, T-duality combined with isometric coordinate shifts 
locally
maps solutions of the original equations of motion to solutions of the
dual equations and thus should preserve the property of
integrability.  Global issues (like twists of boundary conditions
which we have encountered in our discussion) need 
special  treatment.

Assuming that integrability survives at the quantum string level
(cf. \cite{BERK}), our observations   suggest that the 
full $\beta$-deformed $\N=4$
SYM theory is integrable for real $\beta$. This is 
supported by
the integrability of the 1-loop gauge-theory dilatation 
operator, i.e.
the integrability in  the {\it small} $\l$ region.

Special examples in which this behavior is clearly realized include
the case of  rational $\beta$-parameter,  or, more generally, theories
related to ${\cal N}=4$ SYM by some sort of orbifold construction.
In the  orbifold construction   some world
sheet fields may  have twisted boundary conditions.
  Due to its
relation to string theory in $AdS_5\times S^5$, the 
corresponding world sheet sigma
model is classically integrable,  and, quite likely,
 integrable to all
orders in the world sheet perturbation theory
(but still at vanishing string coupling $g_s$).
On the gauge theory side,  the dilatation operator has the same local
structure as in the undeformed theory and is modified only due
to ``boundary effects''\cite{ide}. The single-trace
gauge-invariant operators in the orbifolded theory are obtained from
those in the  original theory by inserting inside the trace the
twist operators representing the orbifold action on the gauge degrees of
freedom.  As a result, the associated dilatation
operator can be represented  by the same spin chain
as in the original theory but with twisted boundary conditions.
Then the relation between the gauge and the string theory integrable
systems is guaranteed (to all orders in
perturbation theory) through the orbifold construction by the
relation  between the parent theories.

A question suggested by  these observations
is whether there exist other transformations which act nontrivially on
sigma models while preserving integrability. Classically,
 any symmetry
of the space of solutions of the classical equations 
of motion of an
integrable theory has this property. Finding such 
 symmetries,
however, is not always straightforward.

For complex deformation parameter $\beta$ the existence of
integrability is not clear;
while the 2-scalar field sector enjoys 1-loop integrability for 
such a
deformation, we have mentioned 
 that at two loops integrability appears to be 
problematic. In the larger 3-spin sector  
 already the 1-loop integrability appears to
be absent \cite{RR,BECH}. On the string side, the lack of
integrability may be related to the use of $S$-duality in the
construction 
of the corresponding background.
It would  be important  to
make this precise and clarify  further 
the lack of
integrability in the presence of a complex deformation.

An interesting extension of our work is to 
the 3-spin sector.
While the string sigma model is likely not to be
integrable  for complex deformation $\beta$, 
it is  still  possible to construct the ``fast
string'' expansion and 
 the analog of the Landau-Lifshitz action.
On the gauge theory side one can also find the coherent state
action for the corresponding ({\it non}-integrable)
 spin chain and compare the two
results. The limit of real
deformation (the case in which all the details should be 
fairly similar to
the ones described in this paper) should 
help to understand how the  integrability
is restored. 
It is possible 
that 
while both the string theory
and the gauge theory lack integrability, 
the associated ``semiclassical'' 
 effective actions describing 
states with  large quantum numbers continue to be the same.
In the limit of vanishing $\beta$-deformation one 
should recover the effective action describing the 
$su(3)$ sector of ${\cal N}=4$ SYM  analyzed  in
\cite{hl,st}.

Another  obvious generalization
is   the analysis of  the $sl(2)$ sector of operators
$\Tr D^S_+ \Phi^J + ...$. Given the way 
the deformation
parameter enters the gauge theory action \rf{vv},
 it is not hard to see that
the  1-loop   dilatation operator in this sector 
should be the same as in
the undeformed case \ci{BEST}. On  the string theory
 side the same is   obviously true
for the leading term of the reduced string
sigma model  action.  
The $AdS_5$ part of the metric is the same, and we
have also an $S^1$ part of the 5-space (the functions
$H$ and $G$ appearing in the deformed solution 
\rf{metrgen} depend only on the
angles of the 5-sphere and are constant 
when restricted to configurations from  the $sl(2)$ sector).
Therefore, the spin chain Landau-Lifshitz action matches 
the string action
as in the undeformed case \cite{st,sl}. At higher loops,
 the situation is less clear and should  be quite interesting
 to study.
  On the string
theory side,  the world sheet sigma model appears unmodified
 at the
classical level, but quantum corrections will feel the deformation 
parameter. 
That suggests 
that on the
 gauge theory side the 
 dependence on the deformation parameter should drop out of
 the 2-loop 
anomalous dimensions to leading order in $1/J$ but 
 should reappear
at subleading orders.

A  more speculative direction is to try to 
reverse-engineer (parts of) supergravity duals of gauge 
theories
exhibiting integrability by starting from 
 the spin chain description
of the dilatation operator.
 In the cases when it is possible to
construct its coherent state continuum limit,  
the resulting action can
 be  interpreted as the ``fast string'' expansion 
 of a sigma model
in the geometry we are interested in constructing.
 The dependence of this
action on the large quantum number may  hint at 
where the fast coordinate
should be reinstated. 
While it is unlikely to be  possible to undo 
 the ``fast string'' limit,
  the result might  in
principle be used as a
 starting point for an ansatz for the complete
solution.  
 It would be interesting to
construct the  supergravity dual of the second exactly marginal
deformation  of  ${\cal N}=4$ SYM. Another apparently simple
deformation on the gauge theory side -- which also breaks conformal
invariance --  is relaxing the relation between
the coefficient of the superpotential and the gauge coupling. This
will produce an ${\cal N}=1$ flow which would 
 be interesting
to study from a supergravity perspective.

{}From a broader perspective, it is hard to underestimate 
the importance
of studying deformations of the $AdS_5\times S^5$ -- ${\cal N}=4$ SYM
duality. In the past, such efforts led to important 
advances  in
our understanding of 4-dimensional field theories. A further 
requirement 
that the deformations preserve a (somewhat fragile) integrable
structure may lead to finer probes of the duality and,  ultimately,  to a
better understanding of the spectrum of strings in $AdS_5\times S^5$
and less-supersymmetric  backgrounds.

\section*{Acknowledgments }

We are grateful to N.~Beisert, I.~Klebanov,   O.~Lunin
  and J.~Maldacena
for useful discussions.
We also thank D.Z. Freedman  for pointing out an error 
in the original version of this paper. 
R.R. is grateful to the Fermi Institute for hospitality
during the initial stages of this work.  His research research was
supported by the DOE grant No.~DE-FG02-91ER40671.
The work of A.A.T.  was supported  by the DOE grant DE-FG02-91ER40690
and also by the INTAS contract 03-51-6346 and the RS Wolfson award.



\appendix


\section{The general spin  chain}
\renewcommand{\theequation}{A.\arabic{equation}}
\setcounter{equation}{0}

There exists a relatively large class of supersymmetric
deformations of ${\cal N}=4$ SYM whose dilatation operators are
described by integrable spin chains.

The Hamiltonian of a spin chain which includes as  special cases
the 2- and 3-field sectors of the $\beta$-deformed ${\cal N}=4$
SYM is \cite{RR}
\begin{eqnarray}
H &=& \sum_{l=1}^L H_{l,l+1}\, ,\\\nonumber H_{l,l+1}
&=&- \frac{\lambda}{8\pi^2}\,{\cal F}\, q\left[ \frac{1-\Upsilon^2
q^2}{1-q^2} \sum_{i}e_{l}^{ii}\otimes e_{l+1}^{ii} +
\frac{q(1-\Upsilon^2)}{1-q^2} \sum_{i\ne j}e^{\alpha_{ij}}
e_{l}^{ij}\otimes e_{l+1}^{ji} \right.\nonumber\\
 && ~~~~~~~~+\left.
\Upsilon^2 \sum_{i< j}e_{l}^{ii}\otimes e_{l+1}^{jj} + \sum_{i>
j}e_{l}^{ii}\otimes e_{l+1}^{jj} \right] \label{sch} \ , 
\end{eqnarray}
where  $L=J$ is the length of the spin chain.\footnote{The
parameter $\Upsilon$ was denoted by $\Delta$ in \cite{RR}. Here we
changed the notation to avoid possible confusion with the standard
notation for conformal dimension.}
 In this expression the indices
$i$ and $j$ take as many values as the number of states $P$ at  each
site of the chain,  and $e_l^{ij}$ are the generators of
the $GL(P)$\  $(e^{ij}_l)_{pq}=\delta^i_p\delta^j_q$ acting at site
$l$.
$\alpha_{ij}$ is an antisymmetric $P \times P$ 
matrix.
For the case of $P=2$  discussed below 
it has single component denoted by $ \alpha$.

\

In \S\ref{spin_chain_stuff} we were  interested in the 2-field
sector of the $\beta$-deformed ${\cal N}=4$ SYM. 
In this case 
the Hamiltonian (\ref{sch}) can be put in a more familiar 
form using
the Pauli matrices
\begin{equation}
e^{12}=\bfsigma^+\ ,  ~~~~~~~e^{21}=\bfsigma^-\ ,  ~~~~~~~
e^{11}=\frac{1}{2}\left(\id+\bfsigma^z\right)\ , ~~~~~~~~
e^{22}=\frac{1}{2}\left(\id-\bfsigma^z\right)~~. \label{pauli}
\end{equation}
The Lagrangian whose 1-loop dilatation operator is the Hamiltonian
(\ref{sch})  (up to the addition of the identity operator and restricted
to the two states per site, $P=2$)
 contains 2 scalar fields $\Phi_i$ in 
 adjoint representation of $SU(N)$
\begin{eqnarray}
{\cal L}=\frac{1}{g^2}\Tr \Bigg[\partial\Phi_i\partial{\bar\Phi}^i\!
&-&\! {\cal F}\,q\,\left[\, k|\Phi_1\Phi_2 - W\,\Phi_2\Phi_1|^2+
C\, (\Phi_1{\bar\Phi}^1+\Phi_2{\bar\Phi}^2)(
{\bar\Phi}^1{\Phi}_1+{\bar\Phi}^2{\Phi}_2)\right.\nonumber\\
&+&\!\left. A \Phi_2\Phi_1{\bar\Phi}^2{\bar\Phi}^1 + B
\Phi_1\Phi_2{\bar\Phi}^1{\bar\Phi}^2 \right]\Bigg] \label{qftlag}
\end{eqnarray}
where
\begin{eqnarray}
\nonumber A=kW+\frac{q(1-\Upsilon^2)}{1-q^2}e^\alpha\ ,  ~~~~
B\!\!&=&\!\!k{\overline
W}+\frac{q(1-\Upsilon^2)}{1-q^2}e^{-\alpha} \ , ~~~~
C=\,\frac{1-\Upsilon^2 q^2}{1- q^2}\ , 
\\
k=-\frac{q^2\,(1-\Upsilon^2)}{1-q^2}\ , &&~~~~~~~~|W|^2=\frac{1}{q^2}\ .
\label{constr1}
\end{eqnarray}
The structure of the last two terms implies that for the theory to
be unitary the parameter $\alpha$ must be chosen to be imaginary.
The Lagrangian (\ref{qftlag}) can be coupled to gauge
field, and  their contribution to (\ref{sch}) is trivial.

In the absence of gauge field, the Lagrangian (\ref{qftlag}) has
four free parameters: the overall coupling constant $g\sqrt{{\cal
F} q k}$, the ratio $C/qk$ and  the real and imaginary parts
of $A/qk$. The equation (\ref{constr1}) puts them in one to one
correspondence with the spin chain parameters $q$, $\Upsilon$ and
$\alpha$. In the presence of gauge field there is an additional
parameter -- the gauge coupling --  which can be identified with
the already present coupling $g$. However, since the gauge
interactions are flavor-blind, it contributes only to rather
trivial shifts in the eigenvalues of the Hamiltonian. Its value
can  be adjusted so that the ground state of $H$ 
has vanishing energy.
 We will use  this additional freedom below.

By adding terms contributing only to nonplanar (and perhaps
self-energy) diagrams, the last term on the first line of \rf{qftlag}
 can be
written as a perfect square of an imaginary combination. Nevertheless,
due to the terms on the second line, the Lagrangian (\ref{qftlag})
cannot be supersymmetrized as long as $A\ne 0$ and $B\ne 0$.

\

The potential term in the Lagrangian (\ref{qftlag}) is  the same
as the potential of the {\it nonconformal} $\beta$-deformed ${\cal
N}=4$ SYM (\ref{vv}) in the 2-field ($\Phi_3=0$) sector
\begin{eqnarray}
V&=&|he^{i\pi\beta}|^2\  \Tr \big[|\Phi_1\Phi_2 -{\bf
q}\Phi_2\Phi_1|^2 \big]-2\Tr\big[~(\Phi_1{\bar
\Phi}^1+\Phi_2{\bar\Phi}^2)({\bar \Phi}^1 {\Phi}_1+ {\bar\Phi}^2{
\Phi}_2)\big] ~~, \label{2fs}
\end{eqnarray}
provided  we make the following identifications (using the
notation in (\ref{not0}) and (\ref{vv}))
\begin{eqnarray}
&&~~~~~~ A=0\ , ~~~~B=0\  ~~~~\Rightarrow~~~~~W=\frac{e^{-2\pi
i\beta_d}}{q} \ , ~~~~~\alpha=-2\pi i\beta_d \ , ~~~~\beta_d\in\Re
\label{constraints_2}\\
&& {\bf q}=W ~~;~~ q=\frac{1}{q_d}=e^{2\pi \kappa_d} ~~;~~
q\,k\,{\cal F}=|he^{i\pi\beta}|^2 ~~;~~ -\frac{C}{k}=
\frac{1-\Upsilon^2q^2}{q^2(1-\Upsilon^2)}
=\frac{2}{|he^{i\pi\beta}|^2}\ . 
\nonumber
\end{eqnarray}
Solving these  conditions   implies that the
 coefficients appearing in this special case of (\ref{sch})
are:
\begin{equation}
\Upsilon^2=\frac{2 -q_d^2|he^{i\pi\beta}|^2}{2-|he^{i\pi\beta}|^2}
\ , ~~~~~~ \frac{1-\Upsilon^2
q^2}{1-q^2}=\frac{2}{2-|he^{i\pi\beta}|^2}\ ,  ~~~~~~
\frac{q^2(1-\Upsilon^2)}{1-q^2}=\frac{|he^{i\pi\beta}|^2}
{2-|he^{i\pi\beta}|^2}\ , 
\label{sol}
\end{equation}
Then, expressing (\ref{sch}) in terms of the Pauli matrices
(\ref{pauli}) and using the freedom of adding the identity operator
to cancel the vacuum energy, the 2-spin interaction Hamiltonian is 
easily found to be 
\begin{eqnarray}
H_2
&=&-\frac{|h|^2\lambda}{16\pi^2} \Bigg[~ \cos2\pi\beta_d\,
(\bfsigma^x\otimes\bfsigma^x + \bfsigma^y\otimes\bfsigma^y) +
\cosh2\pi\kappa_d\,
\left(\bfsigma^z\otimes\bfsigma^z-\id\otimes\id\right)\cr &&
\hphantom{-\frac{\lambda}{16\pi^2} } +\sin2\pi\beta_d \,
(\bfsigma^x\otimes\bfsigma^y-\bfsigma^y\otimes\bfsigma^x) +
\sinh2\pi\kappa_d\,(\bfsigma^z\otimes 1- 1\otimes\bfsigma^z)
\Bigg]\ .  \label{2site}
\end{eqnarray}
Summing over all sites and requiring periodic boundary conditions, 
the last term in \rf{2site} 
disappears and we find that the Hamiltonian of the
spin chain\footnote{ It turns out that this chain has  the
affine quantum symmetry $\widehat{SU(2)}_q$. In \cite{ABRI} this 
Hamiltonian was constructed in the standard way {}from the
generators of this algebra. Translating to our notation, $q_d$ is
related to the quantum deformation while $\beta_d$ is related to
the central extension.} describing the critical $\beta$-deformed
theory is
\begin{eqnarray} \nonumber 
 H&=&\!-\frac{|h|^2\lambda}{16\pi^2}
\sum_{l=1}^L \Bigg[ \hphantom{+}
\cosh2\pi\kappa_d\,\left(\bfsigma^z_l\otimes\bfsigma^z_{l+1}
-\id_l\otimes\id_{l+1}\right)\\
&+&\! \cos2\pi\beta_d\, (\bfsigma^x_l\otimes\bfsigma^x_{l+1} +
\bfsigma^y_l\otimes\bfsigma^y_{l+1} )+ \sin2\pi\beta_d \,
(\bfsigma_l^x\otimes\bfsigma_{l+1}^y-\bfsigma_l^y\otimes
\bfsigma_{l+1}^x) \Bigg]\ . \label{hamiltonian} 
\end{eqnarray}
This is the Hamiltonian for an XXZ spin chain with broken parity
invariance.

The 3-field sector of the $\beta$-deformed SYM is described by a
particular case of (\ref{sch}) only if $q_d=1$ \cite{RR}. It was
argued in \cite{BECH} that no integrable spin chain description
exists for the dilatation operator in the presence of a general
deformation $q_d\in \Re$.

\


\renewcommand{\theequation}{B.\arabic{equation}}
\setcounter{equation}{0}

\section{Asymptotics of the monodromy matrix}

Here we shall provide some details about asymptotics
of the monodromy matrix  and quasimomentum
used in section 7.

The Lax connection $\RR_{1}$ is not convenient to
analyze the asymptotic behavior of the quasi-momentum at infinity,
because it does not vanish at large values of the spectral
parameter $x$. To study the asymptotics it is useful to make an
inverse gauge transformation with the matrix $\M^{-1}$, and use a
nonlocal and nonperiodic Lax connection (\ref{Lax}) with the field
$g$ depending on $\ttp_i$ which satisfy the twisted boundary
conditions (\ref{dphisu2}). Since the matrix $\M$ is not periodic,
the monodromy matrix T$(x)$ is not similar to the path-ordered
exponential of the Lax connection (\ref{Lax}) but is related to it
as follows \bea \la{tranT} \T(x) \= \M(2\pi)\, {\cal
P}\exp\int_0^{2\pi}{\rm d}\s~ \A_{1}(x)\, \M^{-1}(0)\, . \eea
Therefore, the quasi-momentum has the following representation in
terms of the twisted Lax connection (\ref{Lax}) \bea \la{quasi2}
&&2\cos p(x) \= \Tr\left( \M_R\cdot {\cal P}\exp\int_0^{2\pi}{\rm
d}\s~
\A_{1}(x)\right)\, ,\\
\nonumber
&&\A_{1}(x)\= {R_1\ov x^2-1} + {x\, R_0\ov x^2-1}\, ,
\eea
where the matrix $\M_R$ is given by
\bea
\la{MR_init}
\M_R \=
\M^{-1}(0)\, \M(2\pi)\=\left(\begin{array}{cc}
 e^{i\pi \g L}& 0 \\
0 & e^{-i\pi \g L}
\end{array}
\right)\, ,\qquad L\equiv J \= J_1 + J_2\, .
\eea
To find the
asymptotic behavior of the quasi-momentum at infinity we expand
the path-ordered exponential in (\ref{quasi2}) in powers of $1/x$
\begin{equation}
\Tr
\left[ \M_R\left( 1 + {i\bs_a\ov x}\int_0^{2\pi}{\rm d}\s R_0^{(a)} +
{i\bs_a\ov x^2}\int_0^{2\pi}{\rm d}\s R_1^{(a)} - {\bs_a\bs_b\ov
x^2}\int_0^{2\pi}{\rm d}\s\int_0^{\s}{\rm d}\s'
R_0^{(a)}(\s)R_0^{(b)}(\s')\+\dots \right)\right] ,
\end{equation}
where we used the representation $R_i = i\bs_a R_i^{(a)}$. Taking into
account that the matrix $\M_R$ is diagonal, we get the following
expression
\bea
2\cos p(x)\= 2\cos(\pi \g L) - {2\ov x}\sin(\pi \g L)\int_0^{2\pi}{\rm
d}\s R_0^{(3)} -{1\ov x^2}\cos(\pi \g L)\left(\int_0^{2\pi}{\rm d}\s
R_0^{(3)}\right)^2 \nonumber &&\\
\nonumber
- {2\ov x^2} \sin(\pi \g L)\left[ \int_0^{2\pi}{\rm d}\s R_1^{(3)}-
\int_0^{2\pi}{\rm d}\s{\rm d}\s'{\mbox{ sign}}(\s-\s')
R_0^{(1)}(\s)R_0^{(2)}(\s')\right] &&\\
\nonumber
-{1\ov x^2} \cos(\pi \g L)\left[ \left(\int_0^{2\pi}{\rm d}\s
R_0^{(1)}\right)^2+\left(\int_0^{2\pi}{\rm d}\s R_0^{(2)}
\right)^2\right] \+\dots \, ,\la{infinity} &&
\eea
Taking into account that (see, e.g. \cite{KMMZ})
\bea \la{R03}
\int_0^{2\pi}{\rm d}\s R_0^{(3)} = 2\pi {2 M \- L\ov\sqrt{\l}}\,
,\qquad M\equiv J_2\, ,
\eea
we find that the first line of (\ref{infinity}) is just an
expansion of
\bea
2\cos\left(  \pi \g L + 2\pi {2 M \- L\ov\sqrt{\l}\, x} \right)\, .
\eea
Since the quasi-momentum is  conserved, the sum of the remaining terms
in (\ref{infinity}) gives a conserved nonlocal integral of motion
$C_\infty(\g)$.

In the undeformed model on $R\times S^3$, $C_\infty(0)$ is equal
to the sum of squares of off-diagonal components of the nonabelian
charge $\int R_0$, and it is convenient to set it to 0, so that a
classical string solution would be dual to an operator {}from the
holomorphic $su(2)$ subsector of $\N =4$ SYM. In fact, in the
undeformed case, one could set the off-diagonal charges to any
values because any string solution with nonvanishing off-diagonal
charges can be transformed to a string solution with the vanishing
off-diagonal charges by using the $SO(4)$ symmetry of $S^3$.

In the $\g$-deformed model the $SO(4)$ symmetry is broken, and the
degeneracy of the undeformed model is lifted, and, therefore,
different values of the integral of motion $C_\infty(\g)$
correspond to nonequivalent solutions of the sigma model on
$R\times S^3_\g$. It is natural to assume that the class of string
solutions dual to operators {}from the holomorphic $su(2)_\g$
subsector of the $\g$-deformed SYM theory is still singled out by
the requirement $C_\infty(\g)=0$.

Imposing this condition, we get the large $x$ asymptotics of the
quasi-momentum
\bea \la{largex_a}
p(x)\=  \pi\g L \- {2\pi(L-2M)\ov
\sqrt{\l} x} \+ \dots \, ,\quad x\to \infty \, .
\eea
The sign was chosen so that to be compatible with the one used in
\cite{KMMZ}.

The asymptotic behavior of the quasi-momentum at zero can be found
by expanding the twisted Lax connection in powers of $x$
\bea \la{expzero}
\pa_1 \- \A_1(x)&\=& \pa_1\+
R_1\+xR_0\+x^2R_1\+\dots\= g^{-1}\left( \pa_1\+ xL_0\+x^2L_1\right)g
\+\dots\\ \nonumber &\ \equiv& g^{-1}\left( \pa_1\-\LL_1(x)\right)g\, ,
\eea
where $L_\a = \pa_\a gg^{-1}$ is the left current. Then the
quasi-momentum $p(x)$ takes the form
\bea
2\cos p(x) \=\Tr\left( \M_R\, g^{-1}(2\pi)\,
{\cal P}\exp\int_0^{2\pi}{\rm d}\s~ \LL_{1}(x)\, g(0) \right)\ &&\nonumber \\
\la{quasizero} =
\Tr\left( \M_L\cdot {\cal P}\exp\int_0^{2\pi}{\rm d}\s~
\LL_{1}(x)\right)\, ,&&
\eea
where the matrix $\M_L$ is given by \cite{SF}
\bea \la{MR}
\M_L \= g(0)\, \M_R\,
g^{-1}(2\pi)\=\left(\begin{array}{cc}
 e^{i\pi \g (L-2M)}& 0 \\
0 & e^{-i\pi \g (L-2M) )}
\end{array}
\right)\, .
\eea
Expanding the path-ordered exponential in powers of $x$, we get
\bea \nonumber
2\cos p(x)&=& 2\cos(\pi \g (L-2M) ) \\
\nonumber
&+& 2 x\sin(\pi \g
(L-2M) )\int_0^{2\pi}{\rm d}\s L_0^{(3)} -x^2\cos(\pi \g
(L-2M))\left(\int_0^{2\pi}{\rm d}\s L_0^{(3)}\right)^2 \\
\nonumber
&-& 2x^2 \sin(\pi \g (L-2M))\left[ -\int_0^{2\pi}{\rm d}\s L_1^{(3)}-
\int_0^{2\pi}{\rm d}\s{\rm d}\s'{\mbox{ sign}}(\s-\s')
L_0^{(1)}(\s)L_0^{(2)}(\s')\right] \\
 &-&x^2 \cos(\pi \g
(L-2M) )\left[ \left(\int_0^{2\pi}{\rm d}\s
L_0^{(1)}\right)^2+\left(\int_0^{2\pi}{\rm d}\s L_0^{(2)}
\right)^2\right] \+\dots \, . \la{zero}
\eea
Taking into account that  \cite{KMMZ}
\bea \la{L03}
\int_0^{2\pi}{\rm d}\s L_0^{(3)} = -2\pi {L\ov\sqrt{\l}}\, ,
\eea
and setting the nonlocal conserved integral of motion given by the sum
of the terms in the second and third lines of (\ref{zero}), we get the
small $x$ asymptotics of $p(x)$
\bea \la{smallx_a}
p(x)\= \pi\g (L-2M) \+ {2\pi L\ov \sqrt{\l} }\, x \- 2\pi m\+ \dots \,
,\quad x\to 0 \, ,
\eea
where $m$ is an integer quasi-momentum number.


\newpage

\begin{thebibliography}{99}

\bibitem{MALD}
   J.M.~Maldacena,
  ``The Large N Limit of Superconformal Field Theories and
   Supergravity'',
   Adv.Theor.Math.Phys. 2 (1998) 231-252;
   Int.J.Theor.Phys. 38 (1999) 1113-1133; [hep-th/9711200].

\bibitem{GKPO}
    S.S. Gubser, I.R. Klebanov, A.M. Polyakov,
    ``Gauge Theory Correlators {}from Non-Critical String Theory'',
     Phys.Lett. B428 (1998) 105-114, [hep-th/9802109].

\bibitem{WITT}
    E.~Witten,
    ``Anti De Sitter Space And Holography'',
    Adv.Theor.Math.Phys. 2 (1998) 253-291, [hep-th/9802150].

\bi{BMN} D.~Berenstein, J.~M.~Maldacena and H.~Nastase, ``Strings
in flat space and pp waves {}from N =4 super Yang Mills,'' JHEP
{\bf 0204}, 013 (2002) [hep-th/0202021].

\bibitem{GKP2}
S.~S.~Gubser, I.~R.~Klebanov and A.~M.~Polyakov,  ``A
semi-classical limit of the gauge/string correspondence,'' Nucl.\
Phys.\ B {\bf 636} (2002) 99,  hep-th/0204051.

\bibitem{FT1}
S.~Frolov and A.~A.~Tseytlin, ``Semiclassical quantization of
rotating superstring in AdS(5) x S(5),'' JHEP {\bf 0206}, 007
(2002) [hep-th/0204226].

\bi{KRU} M.~Kruczenski, ``Spiky strings and single trace operators
in gauge theories,'' hep-th/0410226.
A.~V.~Belitsky, A.~S.~Gorsky and G.~P.~Korchemsky,
  ``Gauge / string duality for QCD conformal operators,''
  Nucl.\ Phys.\ B {\bf 667}, 3 (2003)
  [hep-th/0304028].

\bibitem{FT}
S.~Frolov and A.~A.~Tseytlin, ``Multi-spin string solutions in
\adss,'' Nucl.\ Phys.\ B {\bf 668}, 77 (2003) [hep-th/0304255].
``Quantizing three-spin string solution in \adss,'' JHEP {\bf
0307}, 016 (2003) [hep-th/0306130].
``Rotating string solutions: AdS/CFT duality in non-supersymmetric
sectors,'' Phys.\ Lett.\ B {\bf 570}, 96 (2003) [hep-th/0306143].

\bibitem{BMSZ}
  N.~Beisert, J.~A.~Minahan, M.~Staudacher and K.~Zarembo,
  ``Stringing spins and spinning strings,''
  JHEP {\bf 0309}, 010 (2003)
  [hep-th/0306139].

\bibitem{BFST}
  N.~Beisert, S.~Frolov, M.~Staudacher and A.~A.~Tseytlin,
  ``Precision spectroscopy of AdS/CFT,''
  JHEP {\bf 0310}, 037 (2003)
  [hep-th/0308117].

\bibitem{AFRT}
G.~Arutyunov, S.~Frolov, J.~Russo and A.~A.~Tseytlin, ``Spinning
strings in \adss and integrable systems,'' Nucl.\ Phys.\ B {\bf
671}, 3 (2003) [hep-th/0307191].

\bibitem{ART}
G.~Arutyunov, J.~Russo and A.~A.~Tseytlin, ``Spinning strings in
\adss: New integrable system relations,'' Phys.\ Rev.\ D {\bf 69},
086009 (2004) [hep-th/0311004].

\bibitem{AS}
G.~Arutyunov and M.~Staudacher, ``Matching higher conserved
charges for strings and spins,'' JHEP {\bf 0403}, 004 (2004)
[hep-th/0310182].
%
``Two-loop commuting charges and the string / gauge duality,''
hep-th/0403077.

\bibitem{SEST}
  D.~Serban and M.~Staudacher,
  ``Planar N = 4 gauge theory and the Inozemtsev long range spin chain,''
  JHEP {\bf 0406}, 001 (2004)
  [hep-th/0401057].



\bibitem{KRUC}
  M.~Kruczenski,
  ``Spin chains and string theory,''
  Phys.\ Rev.\ Lett.\  {\bf 93}, 161602 (2004)
  [hep-th/0311203].

\bibitem{KRTS}
  M.~Kruczenski, A.~V.~Ryzhov and A.~A.~Tseytlin,
  ``Large spin limit of AdS(5) x S5 string theory and low energy
expansion of ferromagnetic spin chains,''
  Nucl.\ Phys.\ B {\bf 692}, 3 (2004)
  [hep-th/0403120].

\bibitem{KMMZ}
  V.~A.~Kazakov, A.~Marshakov, J.~A.~Minahan and K.~Zarembo,
  ``Classical / quantum integrability in AdS/CFT,''
  JHEP {\bf 0405}, 024 (2004)
  [hep-th/0402207].

\bi{KZ} V.~A.~Kazakov and K.~Zarembo,
``Classical / quantum
integrability in non-compact sector of AdS/CFT,''
JHEP {\bf 0410}, 060 (2004)
[hep-th/0410105].
%
N.~Beisert, V.~A.~Kazakov and K.~Sakai,  ``Algebraic Curve for the
SO(6) sector of AdS/CFT,''  hep-th/0410253.
S.~Schafer-Nameki, ``The algebraic curve of 1-loop planar N = 4
SYM,'' hep-th/0412254.
N.~Beisert, V.~A.~Kazakov, K.~Sakai and K.~Zarembo, ``The
algebraic curve of classical superstrings on AdS(5) x S5,''
hep-th/0502226.
L.~F.~Alday, G.~Arutyunov and A.~A.~Tseytlin, ``On integrability
of classical superstrings in AdS(5) x S5,''
hep-th/0502240.



\bi{TSE}
A.~A.~Tseytlin,
``Semiclassical strings and AdS/CFT,''
in:  Proceedings of NATO Advanced Study Institute
and EC Summer School on String Theory: {}from
Gauge Interactions to Cosmology,
Cargese, France, 7-19 Jun 2004.
hep-th/0409296.


\bibitem{MZ}
 J.~A.~Minahan and K.~Zarembo,
 ``The Bethe-ansatz for N = 4 super Yang-Mills,''
 JHEP {\bf 0303}, 013 (2003)
 [hep-th/0212208].

\bi{BKS}
N.~Beisert, C.~Kristjansen and M.~Staudacher,
``The dilatation operator of N = 4 super Yang-Mills
theory,''
Nucl.\ Phys.\ B {\bf 664}, 131 (2003)
[hep-th/0303060].


\bibitem{BEST}
N.~Beisert, {``The complete one-loop dilatation operator of N
= 4 super Yang-Mills theory,''} Nucl.\ Phys.\ B {\bf 676} (2004) 3, {\tt hep-th/0307015};
%
N.~Beisert and M.~Staudacher, {``The N = 4 SYM integrable
super spin chain,''} Nucl.\ Phys.\ B {\bf 670} (2003) 439, {\tt
hep-th/0307042}.


\bibitem{Be}
N.~Beisert, {``Higher loops, integrability and the near BMN
limit,''} JHEP {\bf 0309} (2003) 062, {\tt hep-th/0308074};
%
 { ``The su(2$|$3) dynamic spin chain,''} Nucl.\
Phys.\ B {\bf 682} (2004) 487, {\tt hep-th/0310252}.


\bibitem{BDS}
N.~Beisert, V.~Dippel and M.~Staudacher, ``A novel long range spin
chain and planar N = 4 super Yang-Mills,'' JHEP {\bf 0407}, 075
(2004) [hep-th/0405001].

\bi{B}
N.~Beisert,
``The dilatation operator of N = 4 super Yang-Mills theory and
integrability,''
Phys.\ Rept.\  {\bf 405}, 1 (2005)
[hep-th/0407277].

\bi{D}
A.~V.~Belitsky, V.~M.~Braun, A.~S.~Gorsky and G.~P.~Korchemsky,
  ``Integrability in QCD and beyond,''
  Int.\ J.\ Mod.\ Phys.\ A {\bf 19}, 4715 (2004)
  [hep-th/0407232].
 A.~V.~Belitsky, S.~E.~Derkachov, G.~P.~Korchemsky and A.~N.~Manashov,
  ``Dilatation operator in (super-)Yang-Mills theories on the light-cone,''
  Nucl.\ Phys.\ B {\bf 708}, 115 (2005)
  [hep-th/0409120].
N.~Beisert, G.~Ferretti, R.~Heise and K.~Zarembo,
  ``One-loop QCD spin chain and its spectrum,''
  hep-th/0412029.
G.~Ferretti, R.~Heise and K.~Zarembo,
  ``New integrable structures in large-N QCD,''
  Phys.\ Rev.\ D {\bf 70}, 074024 (2004)
  [hep-th/0404187].

\bi{BTZ}
N.~Beisert, A.~A.~Tseytlin and K.~Zarembo,
 ``Matching quantum strings to quantum spins: One-loop vs.
finite-size corrections,''
 hep-th/0502173.

\bibitem{AFST}
  G.~Arutyunov, S.~Frolov and M.~Staudacher,
  ``Bethe ansatz for quantum strings,''
  JHEP {\bf 0410}, 016 (2004)
  [hep-th/0406256].

\bibitem{Staud}
M.~Staudacher, ``The factorized S-matrix of CFT/AdS,''
hep-th/0412188.


\bi{KS}
S.~Kachru and E.~Silverstein,
 ``4d conformal theories and strings on orbifolds,''
 Phys.\ Rev.\ Lett.\  {\bf 80}, 4855 (1998)
 [hep-th/9802183].

\bi{ide}
K.~Ideguchi,
``Semiclassical strings on AdS(5) x S5/Z(M) and operators in
orbifold field theories,''
JHEP {\bf 0409}, 008 (2004) [hep-th/0408014].

\bi{old}
A.~Parkes and P.~C.~West,
  ``Finiteness In Rigid Supersymmetric Theories,''
  Phys.\ Lett.\ B {\bf 138}, 99 (1984).
D.~R.~T.~Jones and L.~Mezincescu,
  ``The Chiral Anomaly And A Class Of Two Loop Finite Supersymmetric Gauge
  Theories,''
  Phys.\ Lett.\ B {\bf 138}, 293 (1984).
A.~J.~Parkes and P.~C.~West,
  ``Three Loop Results In Two Loop Finite Supersymmetric Gauge Theories,''
  Nucl.\ Phys.\ B {\bf 256}, 340 (1985).
M.~T.~Grisaru, B.~Milewski and D.~Zanon,
  ``The Structure Of Uv Divergences In Ssym Theories,''
  Phys.\ Lett.\ B {\bf 155}, 357 (1985).
D.~R.~T.~Jones,
  ``Coupling Constant Reparametrization And Finite Field Theories,''
  Nucl.\ Phys.\ B {\bf 277}, 153 (1986).
   I.~Jack, D.~R.~T.~Jones and C.~G.~North,
  ``N = 1 supersymmetry and the three loop gauge beta function,''
  Phys.\ Lett.\ B {\bf 386}, 138 (1996)
  [hep-ph/9606323].

  \bi{za}
   A.~Mauri, S.~Penati, A.~Santambrogio and D.~Zanon,
  ``Exact results in planar N = 1 superconformal Yang-Mills theory,''
  hep-th/0507282.

  

\bibitem{LEST}
  R.~G.~Leigh and M.~J.~Strassler,
  ``Exactly marginal operators and duality in four-dimensional N=1
  supersymmetric gauge theory,''
  Nucl.\ Phys.\ B {\bf 447}, 95 (1995)
  [hep-th/9503121].

\bibitem{LUMA}
  O.~Lunin and J.~Maldacena,
  ``Deforming field theories with U(1) x U(1) global symmetry and
    their gravity duals,''
  hep-th/0502086.

\bibitem{AKYA}
  O.~Aharony, B.~Kol and S.~Yankielowicz,
  ``On exactly marginal deformations of N = 4 SYM and type IIB
    supergravity on  AdS(5) x S5,''
  JHEP {\bf 0206}, 039 (2002)
  [hep-th/0205090].

\bi{RT}
J.~G.~Russo and A.~A.~Tseytlin,
 ``Exactly solvable string models of curved space-time backgrounds,''
 Nucl.\ Phys.\ B {\bf 449}, 91 (1995)
 [hep-th/9502038].
``Magnetic flux tube models in superstring theory,''
 Nucl.\ Phys.\ B {\bf 461}, 131 (1996)
 [hep-th/9508068].

\bibitem{RR}
  R.~Roiban,
  ``On spin chains and field theories,''
  JHEP {\bf 0409}, 023 (2004)
  [hep-th/0312218].

\bibitem{BECH}
  D.~Berenstein and S.~A.~Cherkis,
  ``Deformations of N = 4 SYM and integrable spin chain models,''
  Nucl.\ Phys.\ B {\bf 702}, 49 (2004)
  [hep-th/0405215].


\bi{TTT}
M.~J.~Duff, H.~Lu and C.~N.~Pope,
  ``AdS(5) x S(5) untwisted,''
  Nucl.\ Phys.\ B {\bf 532}, 181 (1998)
  [hep-th/9803061].
K.~Sfetsos,
  ``Duality and Restoration of Manifest Supersymmetry,''
  Nucl.\ Phys.\ B {\bf 463}, 33 (1996)
  [hep-th/9510034].
A.~A.~Tseytlin,
  ``Closed superstrings in magnetic flux tube background,''
  Nucl.\ Phys.\ Proc.\ Suppl.\  {\bf 49}, 338 (1996)
  [hep-th/9510041].

\bi{KT}
M.~Kruczenski and A.~A.~Tseytlin,
  ``Semiclassical relativistic strings in S5 and long coherent operators in N
  = 4 SYM theory,''
  JHEP {\bf 0409}, 038 (2004)
  [hep-th/0406189].

\bibitem{SF} 
S.~Frolov,
  ``Lax Pair for Strings in Lunin-Maldacena Background,''
 hep-th/0503201.



\bibitem{BELE}
  D.~Berenstein and R.~G.~Leigh,
  ``Discrete torsion, AdS/CFT and duality,''
  JHEP {\bf 0001} (2000) 038
  [hep-th/0001055].

\bibitem{GSclass}
  R.~R.~Metsaev and A.~A.~Tseytlin,
  ``Type IIB superstring action in AdS(5) x S(5) background,''
  Nucl.\ Phys.\ B {\bf 533}, 109 (1998)
  [hep-th/9805028];
  R.~Kallosh, J.~Rahmfeld and A.~Rajaraman,
  ``Near horizon superspace,'' JHEP {\bf 9809}, 002 (1998)
    [hep-th/9805217];
  R.~Roiban and W.~Siegel,
  ``Superstrings on AdS(5) x S(5) supertwistor space,''
  JHEP {\bf 0011}, 024 (2000)
  [hep-th/0010104].


\bibitem{MTTH}
  R.~R.~Metsaev, C.~B.~Thorn and A.~A.~Tseytlin,
  ``Light-cone superstring in AdS space-time,''
  Nucl.\ Phys.\ B {\bf 596}, 151 (2001)
  [hep-th/0009171].

\bi{BU}
T.~H.~Buscher,
 ``Path Integral Derivation Of Quantum Duality In Nonlinear Sigma
Models,''
 Phys.\ Lett.\ B {\bf 201}, 466 (1988).
A.~S.~Schwarz and A.~A.~Tseytlin,
 ``Dilaton shift under duality and torsion of elliptic complex,''
 Nucl.\ Phys.\ B {\bf 399}, 691 (1993)
 [hep-th/9210015].

\bibitem{TDGS}
  S.~F.~Hassan,
  ``T-duality, space-time spinors and R-R fields in curved backgrounds,''
  Nucl.\ Phys.\ B {\bf 568}, 145 (2000)
  [hep-th/9907152].
M.~Cvetic, H.~Lu, C.~N.~Pope and K.~S.~Stelle,
  ``T-duality in the Green-Schwarz formalism, and the massless/massive
    IIA duality map,''
  Nucl.\ Phys.\ B {\bf 573}, 149 (2000)
  [hep-th/9907202].

 \bi{TKR}
 B.~Kulik and R.~Roiban,
  ``T-duality of the Green-Schwarz superstring,''
  JHEP {\bf 0209}, 007 (2002)
  [hep-th/0012010].

\bibitem{MSW}
  G.~Mandal, N.~V.~Suryanarayana and S.~R.~Wadia,
  ``Aspects of semiclassical strings in AdS(5),''
  Phys.\ Lett.\ B {\bf 543}, 81 (2002)
  [hep-th/0206103].


\bi{BPR}
 I.~Bena, J.~Polchinski and R.~Roiban,
 ``Hidden symmetries of the AdS(5) x S5 superstring,''
 Phys.\ Rev.\ D {\bf 69}, 046002 (2004)
 [hep-th/0305116].

\bibitem{VALL}
  B.~C.~Vallilo,
  ``Flat currents in the classical AdS(5) x S5 pure spinor superstring,''
  JHEP {\bf 0403}, 037 (2004)
  [hep-th/0307018].

\bibitem{POLY}
  A.~M.~Polyakov,
  ``Conformal fixed points of unidentified gauge theories,''
  Mod.\ Phys.\ Lett.\ A {\bf 19}, 1649 (2004)
  [hep-th/0405106].


\bibitem{TSDD}
  A.~A.~Tseytlin,
  ``On dilaton dependence of type II superstring action,''
  Class.\ Quant.\ Grav.\  {\bf 13}, L81 (1996)
  [hep-th/9601109].

\bibitem{TSDS}
  A.~A.~Tseytlin,
  ``Self-duality of Born-Infeld action and Dirichlet 3-brane of type IIB
  superstring theory,''
  Nucl.\ Phys.\ B {\bf 469}, 51 (1996)
  [hep-th/9602064].

\bi{MIHA}
A.~Mikhailov,
``Speeding strings,''
JHEP {\bf 0312}, 058 (2003)
[hep-th/0311019].
``Slow evolution of nearly-degenerate extremal surfaces,''
hep-th/0402067.

\bi{MINA}
J.~A.~Minahan,
 ``Circular semiclassical string solutions on AdS(5) x S5,''
 Nucl.\ Phys.\ B {\bf 648}, 203 (2003)
 [hep-th/0209047].
J.~Engquist, J.~A.~Minahan and K.~Zarembo,
 ``Yang-Mills duals for semiclassical strings on AdS(5) x S5,''
 JHEP {\bf 0311}, 063 (2003)
 [hep-th/0310188].
J.~A.~Minahan,
 ``Higher loops beyond the SU(2) sector,''
 JHEP {\bf 0410}, 053 (2004)
 [hep-th/0405243].

\bibitem{FADT}
  L.~D.~Faddeev and L.~A.~Takhtajan,
  ``Hamiltonian Methods In The Theory Of Solitons'', Springer-Verlag 1987

\bibitem{Fad}
L.~D.~Faddeev,
``How Algebraic Bethe Ansatz works for integrable model,''
hep-th/9605187.

\bibitem{DVTA}
  P.~Di Vecchia and A.~Tanzini,
  ``N = 2 super Yang-Mills and the XXZ spin chain,''
  hep-th/0405262.

\bi{SA}
S. Sachdev, ``Quantum Phase Transitions'', C.U.P., 1999.
S.~Randjbar-Daemi, A.~Salam and J.~Strathdee,
  ``Generalized spin systems and sigma models,''
  Phys.\ Rev.\ B {\bf 48}, 3190 (1993)
  [hep-th/9210145].

\bibitem{NIPR}
  V.~Niarchos and N.~Prezas,
  ``BMN operators for N = 1 superconformal Yang-Mills theories and
    associated string backgrounds,''
  JHEP {\bf 0306}, 015 (2003)
  [hep-th/0212111].

\bibitem{SAZA}
  A.~Santambrogio and D.~Zanon,
  ``Exact anomalous dimensions of N = 4 Yang-Mills operators with
    large R charge,''
  Phys.\ Lett.\ B {\bf 545}, 425 (2002)
  [hep-th/0206079].

\bi{HLO}
R.~Hernandez, E.~Lopez, A.~Perianez and G.~Sierra,
  ``Finite size effects in ferromagnetic spin chains and quantum
    corrections to classical strings,''
  hep-th/0502188.

\bibitem{INOZ}
  V.~I.~Inozemtsev,
  ``Integrable Heisenberg-van Vleck chains with variable range exchange,''
  Phys.\ Part.\ Nucl.\  {\bf 34}, 166 (2003)
  [Fiz.\ Elem.\ Chast.\ Atom.\ Yadra {\bf 34}, 332 (2003)]
  [hep-th/0201001].

\bibitem{HALD}
  F.~D.~M.~Haldane, Z.~N.~C.~Ha, J.~C.~Talstra, D.~Bernard and
V.~Pasquier,
  ``Yangian symmetry of integrable quantum chains with long range
  interactions
  and a new description of states in conformal field theory,''
  Phys.\ Rev.\ Lett.\  {\bf 69}, 2021 (1992).

\bibitem{lup}
K.~Pohlmeyer, {``Integrable Hamiltonian Systems And
Interactions Through Quadratic Constraints,''} Commun.\ Math.\
Phys.\  {\bf 46}, 207 (1976).


\bibitem{ZM}
V.~E.~Zakharov and A.~V.~Mikhailov, {``Relativistically
Invariant Two-Dimensional Models In Field Theory Integrable By The
Inverse Problem Technique,''} Sov.\ Phys.\ JETP {\bf 47} (1978)
1017, [Zh.\ Eksp.\ Teor.\ Fiz.\  {\bf 74} (1978) 1953].


\bibitem{FR}
L.~D.~Faddeev and N.~Y.~Reshetikhin, {``Integrability Of The
Principal Chiral Field Model In (1+1)-Dimension,''} Annals Phys.\
{\bf 167} (1986) 227.

\bibitem{AF}
G.~Arutyunov and S.~Frolov, ``Integrable Hamiltonian for classical
strings on AdS(5) x S5,'' JHEP {\bf 0502}, 059 (2005)
hep-th/0411089.

\bibitem{RS} N. Reshetikhin and F. Smirnov, {``Quantum Floquet
functions''},
Zapiski nauchnich seminarov LOMI (Notes of scientific seminars of
Leningrad Branch of Steklov Institute), v.{\bf 131} (1993) 128 (in
Russian).

\bibitem{Plefka}
T.~Fischbacher, T.~Klose and J.~Plefka, ``Planar plane-wave matrix
theory at the four loop order: Integrability without BMN
scaling,'' JHEP {\bf 0502}, 039 (2005) [hep-th/0412331].

\bibitem{Callan}
C.~G.~Callan, H.~K.~Lee, T.~McLoughlin, J.~H.~Schwarz, I.~Swanson
and X.~Wu, {``Quantizing string theory in $AdS_5 \times S5$:
Beyond the pp-wave,''} Nucl.\ Phys.\ B {\bf 673} (2003) 3, {\tt
hep-th/0307032};
%
C.~G.~Callan, T.~McLoughlin and I.~Swanson, ``Holography beyond
the Penrose limit,'' Nucl.\ Phys.\ B {\bf 694}, 115 (2004)
[hep-th/0404007].
%
C.~G.~Callan, T.~McLoughlin and I.~Swanson, ``Higher impurity
AdS/CFT correspondence in the near-BMN limit,'' Nucl.\ Phys.\ B
{\bf 700}, 271 (2004) [hep-th/0405153].
%
T.~McLoughlin and I.~Swanson, ``N-impurity superstring spectra
near the pp-wave limit,'' Nucl.\ Phys.\ B {\bf 702}, 86 (2004)
[hep-th/0407240].




\bibitem{ABRI}
  J.~Abad and M.~Rios,
  ``Integrable spin chains associated to sl-q(n) and Sl-p,q(n),''
  J.\ Phys.\ A {\bf 28}, 3319 (1995)
  [hep-th/9410193].

\bibitem{GMRO}
  D.~J.~Gross, A.~Mikhailov and R.~Roiban,
  ``Operators with large R charge in N = 4 Yang-Mills theory,''
  Annals Phys.\  {\bf 301}, 31 (2002)
  [hep-th/0205066].

\bibitem{GRMA}
  M.~P.~Grabowski and P.~Mathieu,
  ``Integrability test for spin chains,''
  J.\ Phys.\ A {\bf 28}, 4777 (1995)
  [hep-th/9412039].

\bibitem{FRPATS}
S.~A.~Frolov, I.~Y.~Park and A.~A.~Tseytlin,
``On one-loop correction to energy of spinning strings in S(5),''
Phys.\ Rev.\ D {\bf 71}, 026006 (2005)
[hep-th/0408187].
I.~Y.~Park, A.~Tirziu and A.~A.~Tseytlin,
``Spinning strings in AdS(5) x S5: One-loop correction to energy in SL(2)
sector,''
hep-th/0501203.


\bi{papa}
G.~Papadopoulos, J.~G.~Russo and A.~A.~Tseytlin,
  ``Solvable model of strings in a time-dependent plane-wave background,''
  Class.\ Quant.\ Grav.\  {\bf 20}, 969 (2003)
  [hep-th/0211289].

\bi{hl}R.~Hernandez and E.~Lopez,
``The SU(3) spin chain sigma model and string theory,'' JHEP {\bf
0404}, 052 (2004) [hep-th/0403139].

\bi{st}B.~J.~Stefanski and A.~A.~Tseytlin, ``Large spin limits of
AdS/CFT and generalized Landau-Lifshitz equations,'' JHEP {\bf 0405},
042 (2004) [hep-th/0404133].

\bi{sl}
S.~Bellucci, P.~Y.~Casteill, J.~F.~Morales and C.~Sochichiu, ``sl(2)
spin chain and spinning strings on AdS(5) x S5,'' Nucl.\ Phys.\ B {\bf
707}, 303 (2005) [hep-th/0409086].


\bibitem{BERK}
  N.~Berkovits,
  ``Quantum consistency of the superstring in AdS(5) x S5 background,''
  hep-th/0411170.







\end{thebibliography}
\end{document}